\documentclass[final,5p,times,floatfix]{elsarticle}

\usepackage{graphicx}
\usepackage[draft]{hyperref}

\newcommand{\mr}[1]{{\mathrm{#1}}} 	

\usepackage{amssymb,color,amsmath,bm,mathrsfs}

\usepackage[T1]{fontenc}
\usepackage{listings}
\def\code{\@ifnextchar[{\@with}{\@without}}%
\def\@with[#1]#2{%
}
\def\@without#1{%
  \section{\protect\detokenize{#1}}%
  \lstinputlisting[]{#1}%
}

\definecolor{greencode}{RGB}{0,128,0}
\definecolor{comment}{RGB}{128,128,128}
\definecolor{orange}{RGB}{255,127,0}
\newcounter{bla}

\usepackage{etoolbox,calc}

\makeatletter
\def\appendixname{Appendix}
\appto\appendix{%
  \addtocontents{toc}{\patch@l@section}
  \appto\appendixname{ }
}
\protected\def\patch@l@section{%
  \patchcmd{\l@section}{1.5em}{\widthof{\appendixname\space}+2.5em}{}{}%
}
\makeatother

\journal{Computer Physics Communications}

\begin{document}

\begin{frontmatter}



\title{ARC 3.0: An expanded Python toolbox for atomic physics calculations}


\author[a]{E. J. Robertson}
\author[b]{N. \v{S}ibali\'c \corref{author}}
\author[a]{R. M. Potvliege}
\author[a]{M. P. A. Jones}

\cortext[author] {Corresponding author.\\\textit{E-mail address:}  nikolasibalic@physics.org}
\address[a]{Joint Quantum Center (JQC) Durham-Newcastle, Department of Physics, Durham University, South Road, Durham, DH1 3LE, United Kingdom}
\address[b]{Universit\'{e} Paris-Saclay, Institut d'Optique Graduate School, CNRS, Laboratoire Charles Fabry, 91127, Palaiseau, France}

\begin{abstract}

ARC 3.0 is a modular, object-oriented Python library combining data and algorithms
to enable the calculation of a range of properties of alkali and divalent
atoms. Building on the initial version of the ARC library
[N. \v{S}ibali\'c \emph{et al.}, Comput. Phys. Commun. \textbf{220}, 319 (2017)],
which focused on
Rydberg states of alkali atoms, this major upgrade introduces support
for divalent atoms. It
also adds new methods for working with atom-surface interactions, for modelling ultracold atoms in optical lattices and for calculating
valence electron wave functions and
dynamic polarisabilities. Such calculations have applications in a variety of fields, e.g., in the quantum simulation of many-body physics, in
atom-based sensing of DC and AC fields (including in microwave and THz metrology) and in the development of quantum gate protocols.
ARC~3.0 comes with an extensive documentation including numerous examples. Its
modular structure facilitates its application to a wide range of problems in atom-based quantum technologies.

\end{abstract}


\end{frontmatter}


{\bf PROGRAM SUMMARY}

\begin{small}
\noindent
{\em Program Title:} ARC 3.0 \\
{\em Licensing provisions:} BSD-3-Clause\\
{\em Programming language:} Python \\
{\em Computer:} i386, x86-64 \\
{\em Operating System:} Linux, Mac OSX, Windows\\
{\em RAM:} of the order of several 100 MB for calculations involving several thousand basis states \\
{\em External Routines:} NumPy \cite{1}, SciPy \cite{1}, Matplotlib  \cite{2}, SymPy \cite{3}, LmFit \cite{4} \\
{\em Nature of problem:} \\ 
The calculation of atomic properties of alkali and divalent atoms including energies, Stark shifts and dipole-dipole interaction strengths using matrix elements evaluated through a variety of means.\\
{\em Solution method:}\\ 
Dipole matrix elements are calculated using an analytical semi-classical approximation or wave functions obtained by numerical integration of the radial Schr\"odinger equation for a one-electron model potential.
Interaction energies and shifts due to external fields are calculated using second order degenerate perturbation theory or exact diagonalisation of the interaction Hamiltonian, yielding results valid even at large external fields or small interatomic separation.\\
{\em Restrictions:}\\
External electric and magnetic field must be parallel to the quantization axis. The accuracy of short range ($\lesssim1~\mu$m) atom~-~atom interaction potentials is limited by the truncation of the basis. Only weak magnetic fields are supported as only linear Zeeman shifts are taken into account.
Calculations for divalent atoms use a single-electron approximation
and calculation of their wave functions is not supported.

\end{small}


\lstset{
  basicstyle=\ttfamily,
  columns=fullflexible,
  keepspaces=true,
}

\section{Introduction}


Neutral atoms are ideal building blocks for both fundamental research and 
technological applications exploiting quantum mechanics.
The stability of their physical properties makes 
them an ideal choice for a sensing medium,
virtually eliminating the need for recalibration~\cite{Budker2007,Meyer2020}. By promoting atoms to highly-excited, long-lived Rydberg states, the sensitivity to applied electric fields can be enhanced by many orders of magnitude compared to the ground state~\cite{Sibalic2018}, which has applications to electric field metrology over a wide range of frequencies  from DC to THz fields~\cite{Meyer2020,Sedlacek2012,Wade2016a,Cox2018,Downes2020}.

 The reproducible properties of 
a given atomic species allows for the creation of hundreds of
identical atomic qubits~\cite{Saffman2016a,ChristianGross2017}. Together with the dramatic increase in the interatomic interaction strength associated with Rydberg excitation, this enables applications in
 quantum simulation~\cite{Saffman2016a, Labuhn2016, Bernien2017,Madjarov2020}
 and quantum
 optics~\cite{Pritchard2010,Dudin2012,Peyronel2012,Busche2017,Adams2020}.


Having precise and accurate values of the relevant atomic parameters (e.g., transition dipole moments, Stark shifts, etc.) is crucial for developing new experiments, comparing to theory, or simply for interpreting the data. 
As the number of potentially relevant parameters vastly exceeds what can be tabulated, there is a need for an atom-calculator computing these parameters on demand.
ARC~1.0~\cite{Sibalic2016d} aimed to provide such a research tool for alkali metals by combining
the best available algorithms and experimental data
into a modular, object-oriented Python library. The open source nature of the
package and the popularity of this programming language
facilitated its rapid adoption by the community  and the inclusion of newly developed methods.


However, a growing number of research groups are now working on Rydberg states of atoms with two valence
electrons \cite{Dunning2016}, such as the alkaline earth elements (e.g., Ca, Sr)
and some of the lanthanides (e.g., Yb, Ho \cite{Hostetter2015}). While several groups
have performed calculations of Stark maps \cite{MiaoChan2001,Millen2011spectro,Ye2013} and
interaction potentials \cite{Vaillant2012,DeSalvo2015,Robicheaux2019},
open-source codes like those provided by ARC~1.0~\cite{Sibalic2016d} and Pairinteraction~\cite{swpi2016} for the alkalis are not available for these species. This motivated the major upgrade
of the ARC library we are presenting here. Specifically,
we have extended ARC to divalent atoms, modifying the algorithms as necessary and including supporting data for
 $^{88}$Sr, $^{40}$Ca and $^{174}$Yb. Methods for calculating long-range
 van der Waals interactions have been extended to \emph{degenerate} perturbative calculations and pair-state calculations to arbitrary \emph{interspecies} calculations.
 This new version of
 ARC also adds functionalities for atom-surface interactions, optical trapping and visualisation of results.
The miniaturisation of atom-based sensors~\cite{Shah2007}, and the strong long-wavelength transitions between Rydberg states~\cite{Sibalic2018},
make atom-surface interaction effects
prominent~\cite{Peyrot2019d}. We have therefore included non-retarded van der Waals
atom-surface potential calculations in ARC 3.0. Data are currently included only for sapphire and perfectly reflective surfaces, but other surfaces can easily be incorporated. 

The importance of optical trapping stimulated the addition of two modules. \texttt{OpticalLattice1D} is dedicated to  optical lattices and allows easy calculation of Bloch bands,
Bloch states and Wannier states. \texttt{DynamicPolarizability} calculates the wavelength-dependent atomic polarizability relevant for optical trapping, and enables searches for ``magic-wavelengths'' where two different atomic states have the same polarizability. 

Finally, to allow easy visualisation of different atomic states, a
module \texttt{Wavefunction} provides sectional views of the atomic wave function for arbitrary atomic states. These are important
both for pedagogical and research purposes,
especially since the size of Rydberg electron
orbitals can be large enough to encompass other atoms, and may even approach the typical length-scale over which external trapping potentials vary \cite{Younge2010,Barredo2019}.

How to start using the library is explained in Section 2.
The rest of the paper gives an overview of the newly implemented calculations, with comments on restrictions and implementation details.

\section{Installation and getting started}

ARC 3.0 is available from the online repository \href{https://pypi.org/}{Python Package Index (PyPI)} and can be
installed from the command line simply by invoking

\noindent\texttt{pip install ARC-Alkali-Rydberg-Calculator}

\noindent This installs the package correct for the user's environment, which can be based on
Python 3 for Windows, Mac OS or Linux.
All the methods discussed in the following can be used after importing
them from the \texttt{arc} library:

\begin{lstlisting}
from arc import *
# write your code
\end{lstlisting}

\noindent The examples of code given in the paper assume that this line
was included at the beginning of the program to import
the library.

The detailed documentation of the ARC library is available online~\cite{arc3_documentation} and is updated at each upgrade of the code.
New users are recommended to consult the accompanying
Jupyter iPython interactive notebooks listed on the ``Getting started with ARC'' page of the online documentation~\cite{arc3_documentation},
which provide examples of  calculations and describe the relevant physics.
Finally, we note that a selection of the functionality of the
ARC library is available at \url{https://atomcalc.jqc.org.uk} as a part of the web-app
Atom Calculator. This web page will also generate code that can be used as
example-on-demand to help users start their own calculations.
Bug reporting, questions and further code development are tracked on the 
project's GitHub page~\cite{arcGitHub}.

\section{Overview of the new functions}

\begin{figure*}[t!]
\begin{center}
\includegraphics[width=\textwidth]{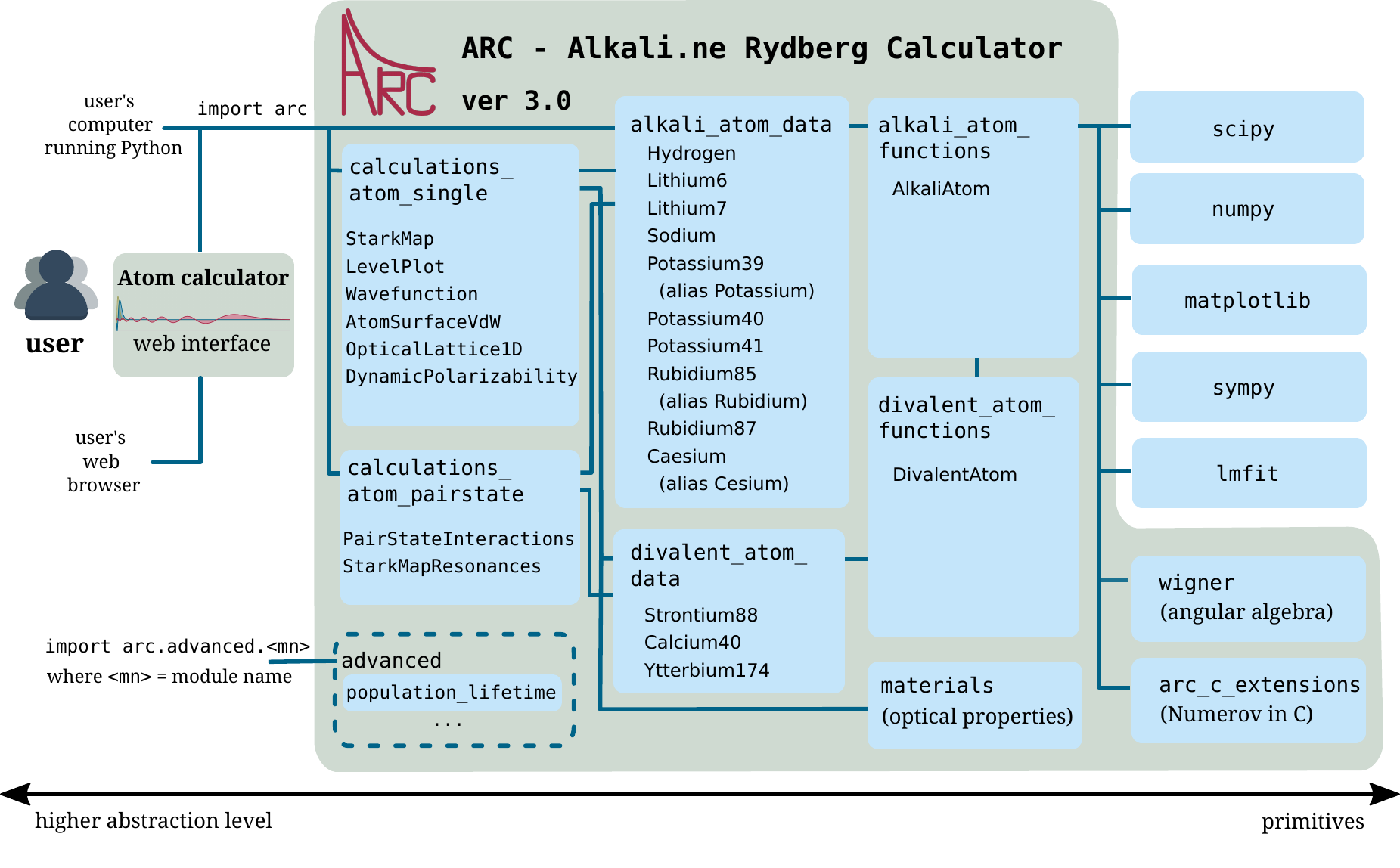}
\end{center}
\caption{\label{fig:module_overview} Overview of ARC~3.0. Only the class names are listed here; a full list of the implemented methods can be found
 in~\ref{app:function_listing}. }
\end{figure*}

A high-level view of the ARC 3.0 library is shown in Fig.~\ref{fig:module_overview}.
This upgrade generalises methods originally included in ARC 1.0, which were developed for \texttt{AlkaliAtom} with electron spin $S=1/2$, to
the singlet ($S=0$) and triplet ($S=1$) states of divalent atoms. 
This approach is underpinned by the broad validity of the single
active electron approximation for highly-excited states of divalent atoms, as discussed in \ref{sec:single_e_approx}.
Thus most of the \texttt{divalent\_atom\_functions}
methods
are directly inherited from \texttt{alkali\_atom\_functions}, upgraded as necessary to support divalent atoms. The spin state is now
specified through an optional keyword
parameter \texttt{s}, which defaults to 0.5 for alkali atoms and must be set to 0 or 1 for divalent atoms. To support these calculations, a semi-classical method for calculating dipole and quadrupole matrix elements has been implemented that does not require the use of a model potential. 

The range of divalent atom data currently included in ARC 3.0 is discussed in Appendix \ref{app:defects}.  New species/series may easily be added by generating the appropriate data files.
Inter-species calculations for both \texttt{PairStateInteractions} and
\texttt{StarkMapResonances} are possible. 

In addition,
\texttt{PairStateInteractions.getC6perturbatively} now supports degenerate perturbation theory for the calculation of $C_6$ coefficients,
The possibility of taking into account a weak magnetic field directed along the quantization
axis has also been introduced (only paramagnetic terms linear in field strength which shift energy levels but do not mix states are included in the calculation).
Moreover, the new single-atom calculations \texttt{Wavefunction}, 
\texttt{AtomSurfaceVdW}, \texttt{OpticalLattice1D} and 
\texttt{DynamicPolarizability} have been added in \linebreak
\texttt{calculations\_atom\_single}.
(The calculations of atom-surface interactions implemented in
\texttt{AtomSurfaceVdW} require values of the refractive index for the surface of interest for a range of 
frequencies. Methods for general optical surface properties are implemented in
\texttt{materials}.)

Finally, we have also created a group of advanced calculations
which are likely to be too specialized for a core toolbox.
The corresponding modules are intended for expert use and can be imported with
\begin{lstlisting}
# from arc.advanced.<mn> import *
# where <mn> is module name, for example:
from arc.advanced.population_lifetime import getPopulationLifetime
\end{lstlisting}
The modules in this advanced library will be built on top of the ARC core library
and will provide solutions for specialised research questions. The first of these modules is \texttt{population\_lifetime}, which
has been contributed by the authors of Ref.~\cite{Archimi2019}. It
gives access to a
\texttt{getPopulationLifetime} 
function calculating {population} lifetimes taking into account the redistribution of population within a Rydberg manifold driven by black-body radiation, including repopulation processes
(such \emph{population} lifetimes thus differ from \emph{state} lifetimes).

\begin{table}[bh!]
\caption{\label{table:data_divalent}
Energy level data for divalent atoms currently included in
\texttt{divalent\_atom\_data}, with references.}
\centering
\begin{tabular}{l|p{2cm} | p{2.5cm} | p{2cm}}
\hline 
Atom & Measured energy levels & Quantum defects & References \\
\hline
$^{88}$Sr & $^1$S$_0$, $^1$P$_1$, $^1$D$_2$, 
$^3$S$_1$, $^3$P$_0$, $^3$P$_1$, $^3$P$_2$,
$^3$D$_1$, $^3$D$_2$, $^3$D$_3$ & 
$^1$S$_0$, $^1$P$_1$, $^1$D$_2$, $^1$F$_3$,
$^3$S$_1$, $^3$P$_0$, $^3$P$_1$, $^3$P$_2$,
$^3$D$_1$, $^3$D$_2$, $^3$D$_3$,
$^3$F$_2$, $^3$F$_3$, $^3$F$_4$,&
\cite{Armstrong1979, Beigang1982a,Beigang1983, Sansonetti2010, Baig1998,Esherick1977,
Esherick1977a, Rubbmark1978, Couturier2019} \\
\hline
$^{40}$Ca &$^1$F$_3$ & $^1$S$_0$, $^1$P$_1$, $^1$F$_3$,
$^3$S$_1$, $^3$P$_1$, $^3$P$_2$, $^3$D$_2$, $^3$D$_1$ & 
\cite{Armstrong1979,Esherick1977,Beigang1982, Gentile1990, Miyabe2006} \\
\hline
$^{174}$Yb &$^1$S$_0$, $^1$P$_1$, $^1$D$_2$, $^3$D$_2$ &
 $^1$S$_0$, $^1$P$_1$, $^1$D$_2$, $^3$D$_2$ &
 \cite{Maeda1992, Aymar1980,  Lehec2018}\\
\hline 
\end{tabular}
\end{table}

In the following we outline the calculations newly implemented in 
ARC 3.0 and provide examples of code for each of them.

\subsection{Divalent atoms dipole and quadrupole matrix elements}\label{sec:semiclassical}

Dipole and quadrupole matrix elements factorize into products of radial matrix
elements and angular factors. 
ARC provides access
to both reduced (e.g. \texttt{getReducedMatrixElementJ}, \texttt{getReducedMatrixElementJ\_asymmetric}, \texttt{getReducedMatrixElementL}, \ldots) and full matrix elements \linebreak
(\texttt{getDipoleMatrixElement}, \texttt{getDipoleMatrixElementHFS}, \ldots)
as detailed in the online documentation~\cite{arc3_documentation}.

As noted above, the calculations are performed within the single active electron approximation. The radial matrix element for a dipole transition between states $|n,L,J\rangle$ and $|n',L',J'\rangle$ of hydrogen or an alkali atom are calculated as in ARC~1.0, i.e., by numerical integration between suitably chosen bounds $r_{\rm i}$ and $r_{\rm o}$:  
\begin{equation}
    \mathcal{R}_{n L J S\, , \,  n'L'J'S'} = \int_{r_{\rm i}}^{r_{\rm o}} R_{nL J}(r)~r~R_{n'L'J'}(r)~r^2~\mathrm{d} r
\end{equation}
with $S=S'=1/2$ and the
wave functions $R_{nL J}(r)$ obtained as solutions of
the Schr{\"o}dinger equation for a model potential.

Since model potential methods are problematic for divalent atoms (\ref{sec:single_e_approx})
the dipole radial matrix elements for these species
are obtained in a semi-classical approximation \cite{Oumarou1988}. This approach is not based on numerical wave functions. Instead, the dipole radial matrix elements take on the form
\begin{equation}
    \mathcal{R}_{nLSJ\, , \, n'L'S'J'} = \frac{3}{2} n_{*\rm c}^2 \left[ 1- \left(\frac{\ell_{\rm c}}{n_{*\rm c}} \right)^2  \right]^{1/2} \sum_{p=0}^{\infty} \gamma^p g_p\left(\Delta n_*\right), 
\end{equation}
with $S=S'=0$ or 1.
In this equation $n^*$ is the reduced principal quantum number ($n_*\equiv n-\delta(n,L,S,J)$ with  $\delta(n,L,S,J)$ being the quantum defect for the $|n,L,S,J\rangle$ state) and
$\ell_{\rm c}$, $n_{*\rm c}$, $\gamma $, $\Delta \ell$ and $\Delta n_*$
are defined as
\begin{align*}
    \ell_{\rm c} &\equiv \frac{L+L'+1}{2}, & n_{*\rm c} &\equiv \sqrt{n'_*\, n_*}, &\gamma &\equiv \Delta \ell \, \ell_{\rm c}/n_{*\rm c}, \\  
    \Delta \ell &\equiv L' - L,& \Delta n_* &\equiv n_* - n'_*.&&
\end{align*}
Moreover, 
\begin{align*}
        g_0(\Delta n_*) &= \frac{1}{3\Delta n_*}\left[\mathcal{J}_{\Delta n_* -1}(-\Delta n_* )-\mathcal{J}_{\Delta n_*+1}(-\Delta n_*)\right],\\
        g_1(\Delta n_*) &= -\frac{1}{3\Delta n_*}\left[\mathcal{J}_{\Delta n_* -1}(-\Delta n_*)+\mathcal{J}_{\Delta n_*+1}(-\Delta n_*)\right],\\
        g_2(\Delta n_*) &= g_0(\Delta n_*) - \frac{\sin{\pi\Delta n_*}}{\pi\Delta n_*},\\
        g_3(\Delta n_*) &= \frac{\Delta n_*}{2}g_0(\Delta n_*)+g_1(\Delta n_*),
\end{align*}{}
where the $\mathcal{J}_s(-\Delta n_*)$'s are Anger functions:
\begin{equation*}
    \mathcal{J}_s(x) \equiv \frac{1}{\pi}\int\limits_0^{\pi} d\theta \cos{\left[ s \theta - x \sin{(\theta)} \right]}.
\end{equation*}

The radial quadrupole matrix elements are calculated as
\begin{equation}
\mathcal{R}^{\rm Q}_{nL JS \, , \, n'L' J'S'} = \int_{r_{\rm i}}^{r_{\rm o}}
R_{n L J}(r)~r^2~R_{n' L' J'}(r)
~r^2~\mr{d} r,
\end{equation}
in the case of hydrogen and alkali atoms, and by using the corresponding semi-classical formulae for divalent atoms.  
The latter differ between
different values of $|\Delta L|$.
For $\Delta L = \pm 2$,
\begin{align}
\mathcal{R}^{\rm Q}_{nLSJ \, , \, n'L'S'J'}
=\frac{5}{2} n_{*\rm c}^{4} &\left [ 1 - \frac{(\ell_{\rm c} +1)^2 }{n_{* \rm c}^{2}}\right ]^{1/2} \nonumber \\ & \times
 \left [ 1 - \frac{(\ell_{\rm c} +2)}{n^{2}_{* \rm c}} \right ]^{1/2} 
 \sum_{p=0}^3 \gamma^p Q_p(\Delta n_*),
\end{align}
whereas for $\Delta L = 0$,
\begin{equation}
\mathcal{R}^{\rm Q}_{nLSJ\, , \, n'L'S'J'} 
= \frac{5}{2} n_{*\rm c}^{4} \left [ 1 - \frac{3\ell_{\rm c}^2}{5n^{2}_{*\rm c}} \sum_{p=0}^1 \gamma^{2p}Q_{2p}(\Delta n_*) \right ].
\end{equation}
The expansion coefficients $Q_p(\Delta n_*)$ are the same in both cases:
\begin{align*}
    Q_0(\Delta n_*) &= -\frac{6}{5(\Delta n_*)^2} g_1(\Delta n_*) , \\
    Q_1(\Delta n_*) &= -\frac{6}{5 \Delta n_*}g_0 (\Delta n_*) + \frac{5}{6} \frac{\sin{\pi\Delta n_*}}{\pi(\Delta n_*)^2}, \\
    Q_2(\Delta n_*) &= -\frac{3}{4}\left[ \frac{6}{5\Delta n_*}g_1(\Delta n_*)+ g_0(\Delta n_*) \right], \\
    Q_3 (\Delta n_*) &= \frac{1}{2}\left[\frac{\Delta n_*}{2} Q_0(\Delta n_*) + Q_1(\Delta n_*) \right].
\end{align*}

We have compared the model potential and semiclassical methods for calculating dipole matrix elements using rubidium as a test case.
The two methods 
give results in close agreement when $|n-n'| \approx 0$, even for $n,n'$ as low as 10 (below which multi-electron effects can be expected to become important). While the agreement deteriorates when $|n-n'|$ increases, the semi-classical results do not differ by more than 5\% from the model potential results in the range $0.65\,n \lesssim n' \lesssim 1.5\,n$. Given that outside this range the dipole matrix elements are less than 1\% of their values at $|n-n'| \approx 0$ (except for very small values of $n$), the semi-classical approach is normally appropriate for any value of $n-n'$ for which the dipole matrix element is large enough to be significant in calculations of Stark maps or dispersion coefficients.
A multi-channel quantum defect approach would be more appropriate in the regions where perturbers mix states of different symmetries~\cite{Vaillant2014}; however, such calculations are beyond the scope of the current version of ARC.

All dipole and quadrupole matrix elements for calcium, strontium and ytterbium currently used or produced by
ARC~3.0 are obtained as described above,
as no literature values of these quantities are currently available
for these atoms. Matrix elements obtained in the future from more accurate
calculations or from measurements can be added to a literature file,
as described in Section~\ref{sec:local_data}.
The library will use the values found in this literature file, should there be any, rather than recalculate them.
   
\subsection{Pair-state calculation of atom-atom $C_6$ interactions in degenerate perturbation theory}

\begin{figure*}[t!]
\begin{center}
\includegraphics[width=\textwidth]{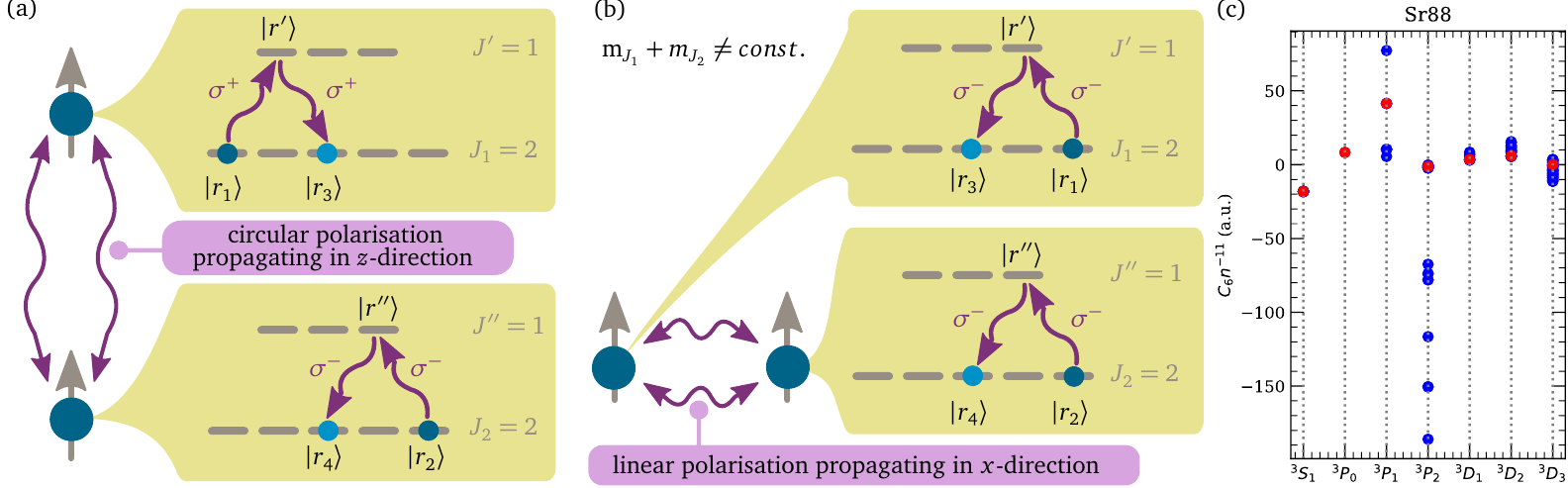}
\end{center}
\caption{\label{fig:c6degenerate} Two examples of cases where degenerate perturbation
theory  has to be used for $C_6$ calculations. (a): Two atoms with inter-atomic axis parallel to the quantisation axis may exchange virtual photons that  drive two $\sigma^+$ transitions
[corresponding to $Y_{1,+1}$ in Eq.~(\ref{eq:Vl1l2})] in one atom
and two $\sigma^-$ transitions in the other atom, thus coupling the initial $|r_1\rangle$ and $|r_2\rangle$ states to energetically permitted
$|r_3\rangle$ and $|r_4\rangle$ states. 
The projection of the total angular momentum on the quantisation axis is conserved in this process (i.e., $m_{J_1} + m_{J_2} = m_{J_3} + m_{J_4}$).
(b): Alternatively, when the inter-atomic axis is
perpendicular to the quantisation axis, virtual photons driving a $\sigma^-$ transition in
one atom may be viewed as having linear polarization perpendicular to the quantisation axis and can thus drive both $\sigma^+$ and $\sigma^-$ transitions
in the other atom. The projection of the total angular momentum on the  quantisation
axis is then not conserved and $m_{J_3}+m_{J_4}$ may differ from $m_{J_1}+m_{J_2}$. Using degenerate perturbation theory, 
the energy eigenstates can be obtained with their corresponding 
$C_6$ coefficients,
as shown in (c) for the case of two $^{88}$Sr atoms initially in a pair state with $n_1=n_2=40$, $L_1=L_2$, $J_1=J_2$ and $S_1=S_2$ (the red circles correspond to the stretched states $m_{J_1}=J_1 = m_{J_2}=J_2$, assuming that
the inter-atomic axis is parallel to the quantization axis).}
\end{figure*}

In the single-active electron approximation, the interaction between two atoms at inter-atomic distance $R$
arises from the interaction between the valence electrons and the interaction of each of these electrons with the screened nucleus of the other atom. Denoting the coordinates of the valence electrons relative to the respective nuclei by $\mathbf{r}_1$
and $\mathbf{r}_2$, this interaction
is given by the following multipolar expansion
(c.f.\ Eq.~(19) in Ref.~\cite{Sibalic2016d} for details):
\begin{eqnarray}
V(R) &=& \sum_{k_1, k_2 = 1}^\infty \frac{V_{k_1,k_2}(\mathbf{r}_1, 
\mathbf{r}_2)}{R^{k_1+k_2+1}},
\end{eqnarray}
with
\begin{eqnarray}
\label{eq:Vgeneral}
V_{k_1,k_2}(\mathbf{r}_1, \mathbf{r}_2)&=& \frac{(-1)^{k_2} 4\pi}{\sqrt{(2k_1+1)(2k_2+1)}} \nonumber \\
& & \times \sum_p \sqrt{\begin{pmatrix}
k_1+k_2 \\
k_1+p
\end{pmatrix}
\begin{pmatrix}
 k_1+k_2 \\
 k_2 +p
\end{pmatrix}
 } \, r_1^{k_1} r_2^{k_2} \nonumber \\
& & \times\,Y_{k_1, p}(\hat{r}_1)  Y_{k_2, -p}(\hat{r}_2).\label{eq:Vl1l2} 
\end{eqnarray}
Here the $\begin{pmatrix}n \\ m \end{pmatrix}$ are binomial coefficients and the
$Y_{k,p}(\hat{r})$ are spherical harmonics, and it is assumed that the quantization axis of both
atoms is directed \emph{along} the internuclear axis
[as in Fig.~\ref{fig:c6degenerate}(a)].
Terms with $k_1+k_2= 2, 3, 4, \ldots$
correspond respectively to dipole-dipole, dipole-quadrupole,
quadrupole-quadrupole interactions and so on.

In the large $R$ limit, the interaction between Rydberg atoms is
typically dominated by dipole-dipole interactions. Such an interaction couples an initial
pair-state $|r_1~r_2\rangle$ to other states $|r'~r''\rangle$ 
whose energy differs by an energy defect $\Delta_{r',r''} = E_{r'}+E_{r''}-E_{r_1}-E_{r_2}$. To second order in this interaction, the interaction energy of two atoms in the initial pair-state $|r_1~r_2\rangle$ 
is given by a van der Waals interaction potential of the form $-C_6/R^6$,
where $C_6$ is defined by the following equation if the pair-state energy $E_{r_1}+E_{r_2}$ is non-degenerate:
\begin{equation}
\frac{C_6}{R^6} = \sum_{r',r''} \frac{|\langle r'~r''\,|\,V_{\rm dd}(R)\,|\,r_1~r_2 \rangle|^2}{\Delta_{r',r''}}.
\end{equation}
Here the sum goes over all the pair states $|r'~r''\rangle$ that are coupled by electric dipole transitions to the 
original pair-state $|r_1~r_2\rangle$, and $V_{\rm dd}(R)$ is the dipole-dipole part of $V(R)$:
\begin{equation}
V_{\rm dd}(R) \equiv V_{1,1}({\bf r}_1,{\bf r}_2)/R^3.
\end{equation}

However, there are situations in which the pair-state energy $E_{r_1}+E_{r_2}$ is degenerate and the dipole-dipole interaction couples the
initial pair-state $|r_1~r_2\rangle$ to some other pair-states $|r_3~r_4\rangle$
of the same
energy --- for instance, in the absence of external fields, to pair-states differing in magnetic quantum numbers only. Two examples of such off-diagonal couplings mixing pair-states are given in Fig.~\ref{fig:c6degenerate}.
In such situations,
a perturbative calculation of the $C_6$ coefficients requires the diagonalization of the matrix
$\mathcal{C}$ describing the second-order coupling between the energy-degenerate pair-states \cite{Vaillant2012}. To this end, we work in the basis $\{|\,r_i~r_j\rangle\}$ of degenerate pair-states, i.e.,
\begin{align}\label{eq:basisC6}
|\,r_i~r_j\rangle &\in \{|\,n_1 L_1 S_1 J_1 m_{J_1}=-J_1\rangle, \ldots , |\,n_1 L_1 S_1 J_1 m_{J_1}=+J_1 \rangle\}\nonumber \\
&\otimes 
\{|\,n_2 L_2 S_2 J_2 m_{J_2}=-J_2\rangle, \ldots , |\,n_2 L_2 S_2 J_2 m_{J_2}=+J_2 \rangle\},
\end{align}
where $(n_1,L_1,S_1,J_1,m_{J_1})$ and $(n_2,L_2,S_2,J_2,m_{J_2})$ are the quantum numbers
associated with the states $|\,r_1\rangle$ of atom~1 and $|\,r_2\rangle$ of atom~2, respectively.
The element of $\mathcal{C}$ corresponding to the second-order coupling between the $q$-th and the $p$-th basis states is
\begin{displaymath}
R^6\,\sum_{r',r''}\frac{\langle r_{i_q}~r_{j_q}\,|\,V_{\rm dd}(R)\,|\,r'~r''\rangle \langle r'~r''\,|\,V_{\rm dd}(R)\, |\, r_{i_p}~r_{j_p}\rangle}{\Delta_{r'r''}}.
\end{displaymath}
Diagonalising this matrix yields a $C_6$ coefficient for each of the energy curves the degenerate pair-state energy splits into due to the dipole-dipole interaction, these coefficients being the eigenvalues $C_6^{(i)}$ of $\mathcal{C}$
[see, e.g., Fig.~\ref{fig:c6degenerate}(c)].
These eigenvalues are \emph{independent} of the orientation of the inter-atomic axis relative to the quantization axis, in the absence of external fields, although the composition of the corresponding energy eigenstates in terms of the basis states defined above depends on the choice of the quantization axis \cite{Vaillant2012}.

Degenerate perturbation theory can be used within the \linebreak \texttt{getC6perturbatively} function by setting the flag
\texttt{degeneratePerturbation=True}. The function  will then return the $C_6^{(i)}$ eigenvalues and the corresponding eigenvectors of the relevant $\mathcal{C}$ matrix.
For example, the following fragment of code produces the results displayed in the third column of Fig.~\ref{fig:c6degenerate}(c)
\begin{lstlisting}
# Sr 88  40 ^3P_1 mj=0 , 40 ^3P_1 mj=0 state
calc = PairStateInteractions(
    Strontium88(),
    40, 1, 1,
    40, 1, 1, 
    0, 0,
    s=1
    )
theta = 0; phi = 0; deltaN = 5;
deltaE = 30e9;  # in Hz 
c6, eigenvectors = calc.getC6perturbatively(
    theta, phi, 5,
    deltaE, degeneratePerturbation=True)
# getC6perturbatively returns the C6 coefficients
# expressed in units of GHz mum^6.
# Conversion to atomic units:
c6 = c6/1.445e-19
# These results should still be divided by n^{11} 
# to be plotted as in Fig. 2(c).
\end{lstlisting}
Here \texttt{theta} and \texttt{phi} are, respectively, the polar angle ($\Theta$) and azimuthal angle ($\Phi$) defining the orientation of the inter-atomic axis in a referential whose $z$-axis is parallel to the axis of quantisation of the angular momenta.
The dependence on $\Theta$ and $\Phi$ of the elements of $\mathcal{C}$ is taken into account
by rotating the atomic basis states using Wigner D-matrices, as in
ARC~1.0~\cite{Sibalic2016d}. I.e., $V_{\rm dd}(R)$ defined above is replaced by the angle-dependent $V_{\rm dd}(R,\Theta,\Phi)$, with, in a simplified notation,
\begin{align}
V_{\rm dd}(R, \Theta,\Phi) = &[\mathrm{D}(J_1',\Theta,\Phi)\otimes \mathrm{D}(J_2',\Theta,\Phi)] \nonumber\\
&V_{\rm dd}(R) \nonumber \\
&[\mathrm{D}(J_1,\Theta,\Phi)\otimes \mathrm{D}(J_2,\Theta,\Phi)]^\dagger,
\end{align}
where the $\mathrm{D}(J\,m_{J},\Theta,\Phi)$ represent the relevant rotation matrices.
Although the elements of the matrix $\mathcal{C}$ and the composition of its eigenstates in terms of the basis states defined above depend on $\Theta$ and $\Phi$, this is not the case for
its eigenvalues $C_6^{(i)}$ (hence for the $C_6$ coefficients resulting from this calculation) \cite{Vaillant2012}.

Invoking \texttt{getC6perturbatively} without the flag \texttt{degeneratePerturbation=True} or with \texttt{degeneratePerturbation=False} will only return the individual element of the matrix $\mathcal{C}$ corresponding to the values of the quantum numbers specified in the call, rather than the eigenvalues and eigenvectors of $\mathcal{C}$. These individual elements normally depend on $\Theta$ and $\Phi$ and can be taken to be effective $C_6$ coefficients for particular combinations of magnetic quantum numbers.

\begin{figure}[!t]
\begin{center}
\includegraphics[width=\linewidth]{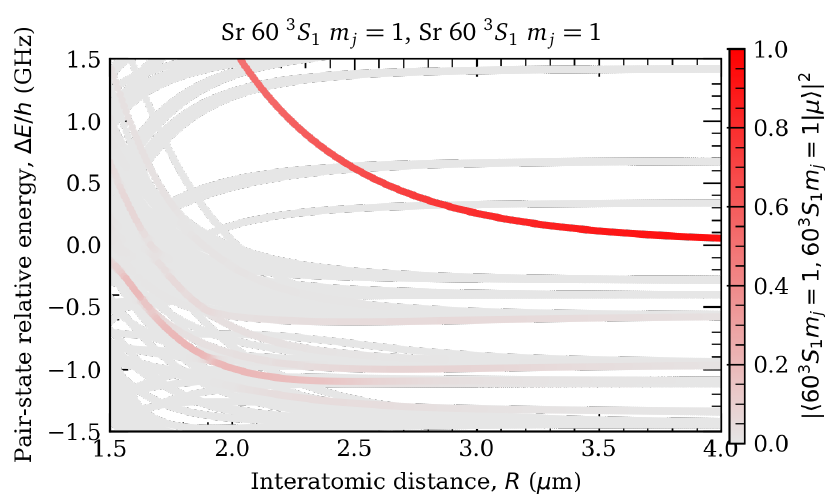}
\end{center}
\vspace{-0.3in}
\caption{\label{fig:pair_state_full}
Example interaction strength calculations using the full diagonalisation method for a pair of divalent atoms in identical states ($^{88}$Sr 5s60s $^3$S$_1 m_{J}=1$). 
The plot shows how the energy levels of the atom pair vary as functions of the inter-atomic distance, $R$. The energies are measured relative to the energy of the initial pair-state in the $R\rightarrow \infty$ limit, and only pair-states coupled to this initial pair-state are represented. 
The fraction of the initial pair state present in each eigenstate is colour-coded as per the colour bar.}
\end{figure}

We note that the interaction energies can also be obtained non-perturbatively by full diagonalisation of the Hamiltonian using the function \texttt{diagonalise}. An example of results obtained in this way is given by Fig.~\ref{fig:pair_state_full}, which is produced by running the following code:
\begin{lstlisting}
calc = PairStateInteractions(
    Strontium88(),
    60, 0, 1,
    60, 0, 1, 
    1, 1,
    s=1
    )
theta=0; phi=0; deltaN = 5; 
deltaL = 5;  deltaMax = 25e9 #  [Hz]

# Generate pair-state interaction Hamiltonian
calc.defineBasis(theta, phi, deltaN, deltaL, deltaMax,
                 progressOutput=True)

# Diagonalise
r = np.linspace(1.5, 4, 300)
nEig = 200  # Number of eigenstates to extract
calc.diagonalise(r, nEig, progressOutput=True)

# Plot    
calc.plotLevelDiagram()
calc.showPlot(interactive=False)
\end{lstlisting}

We also note that the implementation of degenerate perturbation theory made in ARC does not take into account energy degeneracies between states differing in $L$ or $S$. It is therefore not appropriate for atomic hydrogen or for high angular momentum states. We recommend using the full diagonalisation method for such cases.

\subsection{Inter-species interaction calculations}

\begin{figure}[!b]
\begin{center}
\includegraphics[width=\linewidth]{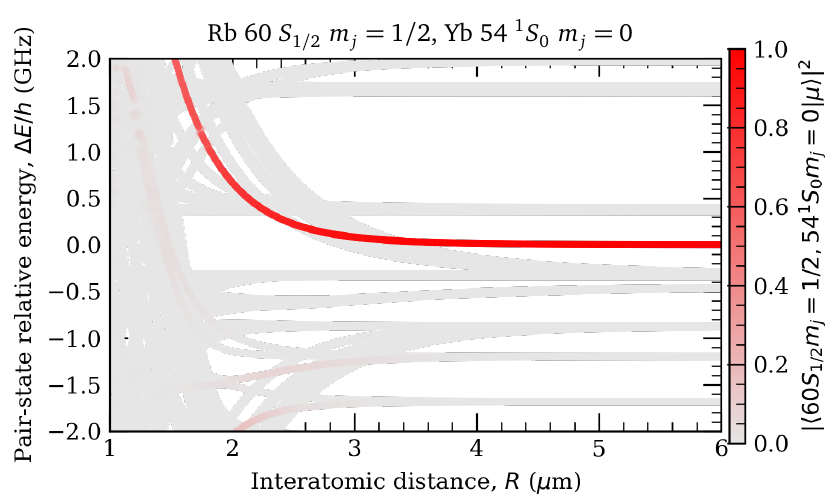}
\end{center}
\vspace{-0.3in}
\caption{\label{fig:pair_state_interspecies}
Example alkali-divalent interspecies interaction calculation ($^{85}$Rb 60s $^2$S$_{1/2} m_{J}=1/2$; $^{174}$Yb 5s54s $^1$S$_{0} m_{J}=0$). The plot shows how the energy levels of the atom pair vary as functions of the inter-atomic distance, $R$. The fraction of the initial pair state present in each eigenstate is colour-coded as per the colour bar.}

\end{figure}

\texttt{PairStateInteractions} supports inter-species calculations.
Users can initialize such calculations using the keyword argument
\texttt{atom2} to explicitly state the species of the second atom. Note that to specify the spin state of the second atom, the keyword argument \texttt{s2} should also be set. Setting \texttt{s2} can also be used for calculations where the atoms are from the same atomic species but have different spin.

For example, pair-state
calculations  between rubidium atoms in the $|\,$5s60s~$^2$S$_{1/2}~m_{J}=1/2\rangle$ state and ytterbium atoms in the
$|\,$6s54s~$^1$S$_0~m_{J}=0\rangle$ state [Fig.~\ref{fig:pair_state_interspecies}] are initialized as follows:
\begin{lstlisting}
calc = PairStateInteractions(
    Rubidium(),
    60, 0, 0.5,
    54, 0, 0, 
    0.5, 0,
    s=0.5,
    atom2=Ytterbium174(),
    s2=0
    )
\end{lstlisting}

\subsection{Single-atom properties for divalent atoms}

\begin{figure}[!t]
\begin{center}
\includegraphics[width=\linewidth]{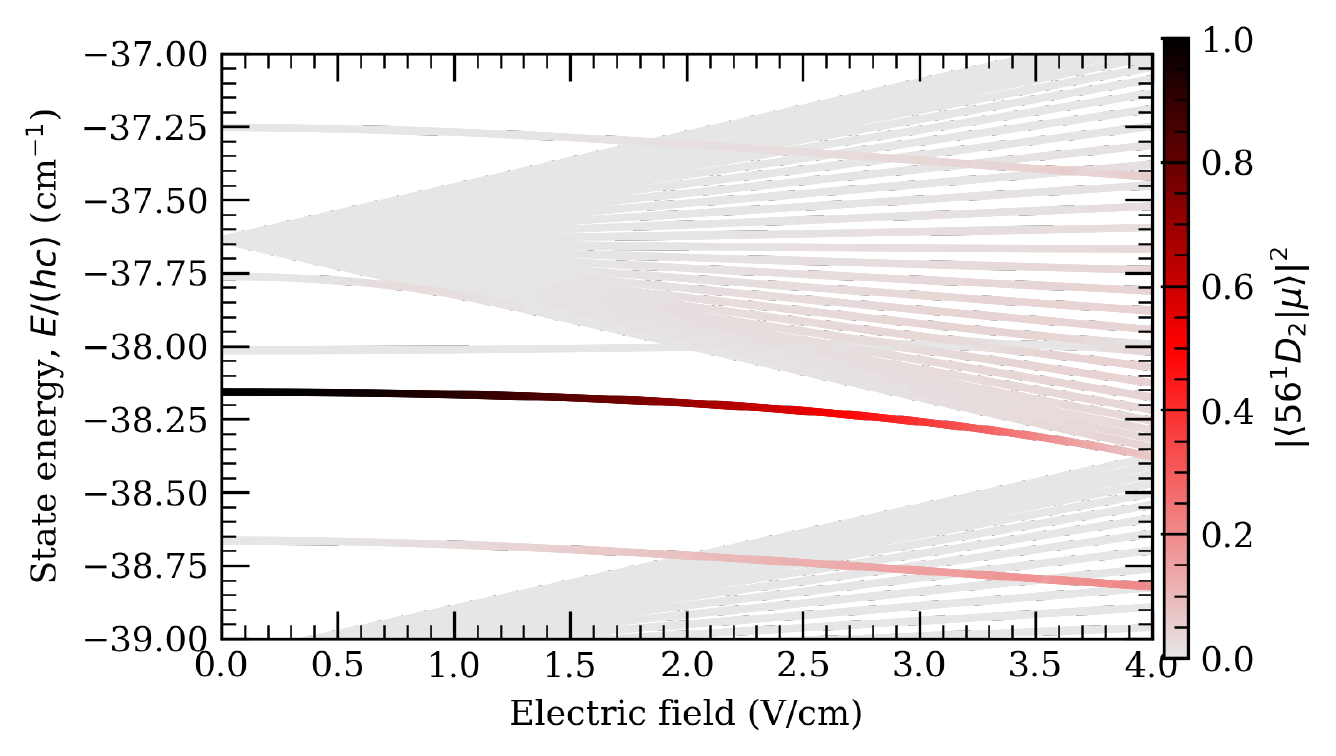}
\end{center}
\vspace{-0.3in}
\caption{\label{fig:stark_Sr} An example Stark map calculation for a divalent atoms,  showing the perturbation of states in the vicinity of the  $|$5s60d~$^1$D$_2$ $m_{J}=0\rangle$ state of $^{88}$Sr by an external electric field.
Only the coupled states (i.e., singlet states with $m_J=0$) are plotted.
The colour bar shows the admixture of this state into each atomic eigenstate.}

\end{figure}

ARC 3.0 extends most of the single atom methods available in ARC 1.0 to divalent atoms. For example, Stark maps can be obtained
for divalent atoms by setting the additional key argument \texttt{s} to define the spin state. Note that in the single electron approximation, states of different total spin are not coupled. Example results from such a calculation are shown in Fig.~\ref{fig:stark_Sr}.
Static electric fields are often used to adjust pair-state energies.
\texttt{StarkMapResonances} allows the user to search for electric field
strengths where two pair-states have same energies (F\"orster resonances).
Lastly \texttt{LevelPlot} allows the plotting and interactive exploration of energy level diagrams. These diagrams may be
opened as interactive stand-alone plots (from a command line Python call
or in a Jupyter notebook with \texttt{\%matplotlib qt}); then will then display the transition wavelength and transition frequency for pairs of states selected
interactively by clicking on energy levels.

\subsection{Electronic wave functions}
Visualisations of atomic wave functions are a useful pedagogical tool enabling visual interpretation of effects such as dipole moments. In addition, the detailed shape of Rydberg electron wave functions plays an important role in a number of effects.
For example, the wave functions can become so spread out that
they encompass other atoms. The modulations of the electron probability density may then
induce a significant variation in the potential energy of the encompassed atoms, which may lead
to the formation of
Rydberg molecules~\cite{Gaj2014}.
At the same time, the optical potentials used for atom trapping may vary substantially
over the length-scale of the Rydberg electron wave functions, giving rise to
energy shifts~\cite{Younge2010} and affecting the trap
lifetime~\cite{Barredo2019}.

\begin{figure}[!t]
\begin{center}
\includegraphics[width=\linewidth]{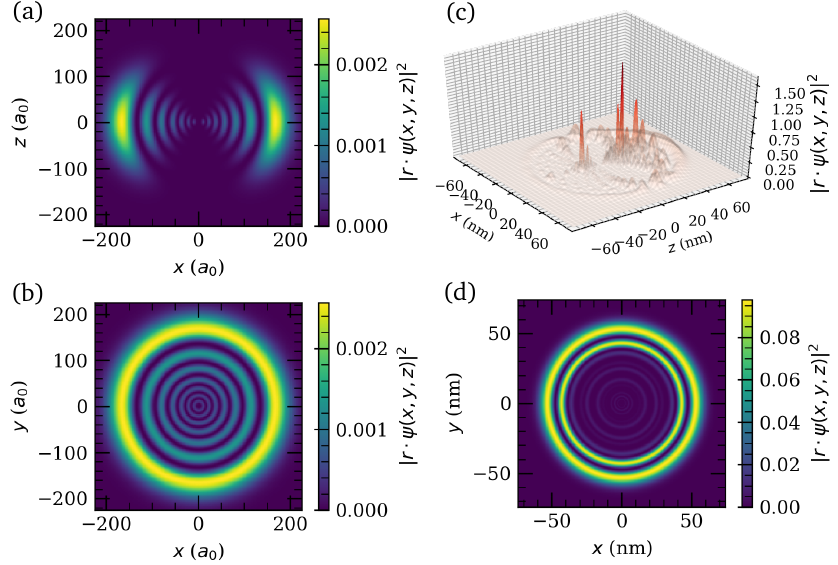}
\end{center}
\vspace{-0.3in}
\caption{\label{fig:wavefunction}
Example wave function visualizations showing  (a and b) cuts through the $^{85}$Rb $|$10f~$^2$F$_{7/2}~m_{J}=7/2\rangle$ state  two orthogonal planes and (c and d) similar cuts for the Cs $|$28s~$^2$S$_{1/2}~m_{J}=1/2\rangle$ state 
perturbed by a strong electric field ($-240$~V/cm). The plot shows the eigenstate that is adiabatically
connected to the $|$28s$~^2$S$_{1/2}~m_{J}=1/2\rangle$ unperturbed state in zero electric field.
Note the change of units ($a_0$ or nm), which can be easily specified using keyword arguments in the \texttt{Wavefunction} methods.}

\end{figure}

\texttt{Wavefunction} enables the calculation of atomic wave functions for arbitrary superposition states. Quick 2D and 3D visualisations are possible, with a  choice of units (atomic or SI).
 For example, the following code can be used to obtain and plot the probability density function
for the 10f~$^2$F$_{7/2}$~$m_J=7/2$ state of rubidium [Figs.~\ref{fig:wavefunction}(a and b)]:

\begin{lstlisting}
atom = Rubidium()
n = 10; l=3; j=3.5;  mj=3.5;
stateBasis = [[n, l, j, mj]]  
stateCoef = [1]  # pure 10 F_7/2 mj=7/2 state
wf = Wavefunction(atom, stateBasis, stateCoef)
wf.plot2D(plane="x-z", units="atomic"); plt.show()
wf.plot2D(plane="x-y", units="atomic"); plt.show()
\end{lstlisting}

\texttt{Wavefunction} can be integrated with other ARC functions. For example, one can find the
electronic wave function for an atom perturbed by an electric field by getting the state from the corresponding
\texttt{StarkMap} and using \texttt{Wavefunction} as per the following code to plot the result. An example is shown in  Figs.~\ref{fig:wavefunction}(c-d).

\begin{lstlisting}
atom = Caesium()
calc = StarkMap(atom)
states, coef, energy= calc.getState([28, 0, 0.5, 0.5], -24000,23,32,20, accountForAmplitude=0.95,
                       debugOutput=True)
wf = Wavefunction(atom, states, coef)
wf.plot3D(plane="x-z", units="nm"); plt.show()
wf.plot2D(plane="x-z", units="nm", pointsPerAxis=400, axisLength=2800)
plt.show()
\end{lstlisting}

The density and scale of the mesh on which the  wave functions are calculated can be adjusted
with optional keyword parameters.
The probability density functions can be provided in Cartesian as well as in spherical coordinates (respectively through 
\texttt{getRtimesPsi} and \texttt{getRtimesPsiSpherical}), and it is also possible to obtain arrays of wave functions for all different possible spin states (using the
\texttt{getPsi} method).

Note that \texttt{Wavefunction} is currently only supported for species in the \texttt{alkali\_atom\_data} class, as calculations based on model potentials are not currently supported for divalent atoms.

\subsection{Atom-surface interactions: van der Waals potentials}

In the vicinity of a surface, an atomic dipole interacts with its image
in the surface, leading to shifts of the atomic energy levels. For small atom-surface
distances $z<\lambda/(2\pi)$, where $\lambda$ is the wavelength associated with the strongest transition,
the interaction potential $V(r)$  for an atom in the state $a$ is of the non-retarded, van der Waals form~\cite{Fichet1995}
\begin{align}
V_{\rm AS}(z) &= -\frac{1}{4\pi\varepsilon_0}
\sum_b
\frac{n(\omega_{ab})^2 - 1}{n(\omega_{ab})^2 + 1}
\frac{|\mu_{x}^{ab}|^2 + 
|\mu_{y}^{ab}|^2 + 
2 |\mu_{z}^{ab}|^2}{16 z^3} \nonumber
\\
&\equiv -\frac{C_3}{z^3}.
\end{align}
Here $n(\omega_{ab})$ is the frequency dependent refractive index associated
with the surface. The summation covers all the states $b$ dipole coupled to
state $a$.
The corresponding transition frequencies and dipole matrix elements
in the $x$, $y$ and $z$ directions are respectively $\omega_{\rm ab}$, 
$\mu_{x}^{ab}$, $\mu_{y}^{ab}$ and
$\mu_{z}^{ab}$.
The $z$-axis is taken to be perpendicular to
the surface. Note that different states will have different $C_3$
values, which leads to a modification of the atomic transition frequencies near the surface.

To specify the surface material, ARC provides an abstract class
\texttt{materials.OpticalMaterial}, with  a method \texttt{getN}
returning the refractive index $n$ for a
specified wavelength. A subclass \texttt{Sapphire} is provided
as an example. The \texttt{AtomSurfaceVdW} uses information on the optical properties of the surface and the atomic transition frequencies to calculate $C_3$. For example, the following code will return the energy shift with error for the 6s~$^2$S$_{1/2}$ state of caesium in the proximity of a sapphire surface (the result is 1.259(2)~kHz~$\mu$m$^3$):
\begin{lstlisting}
from arc.materials import Sapphire

atom = Cesium()
surface = Sapphire()
calc = AtomSurfaceVdW(atom, surface)

# look at 6 S_1/2 state
n1 = 6; l1 = 0; j1 = 0.5
# take into account coupling to
# 6 P_1/2, 6 P_3/2, 7 P_1/2, 7 P_3/2
coupledStates = [[6, 1, 0.5],
    [6, 1, 1.5],
    [7, 1, 0.5],
    [7, 1, 1.5]]
c3, c3_err = calc.getStateC3(n1, l1, j1,
                             coupledStates,
                             debugOutput=True)
\end{lstlisting}

Such calculations are possible for all the atomic species supported by ARC, to the extent that the required dipole matrix elements are available.

\subsection{Optical lattices: Bloch bands, Bloch states, Wannier states}
\begin{figure}[!b]
\begin{center}
\includegraphics[width=\linewidth]{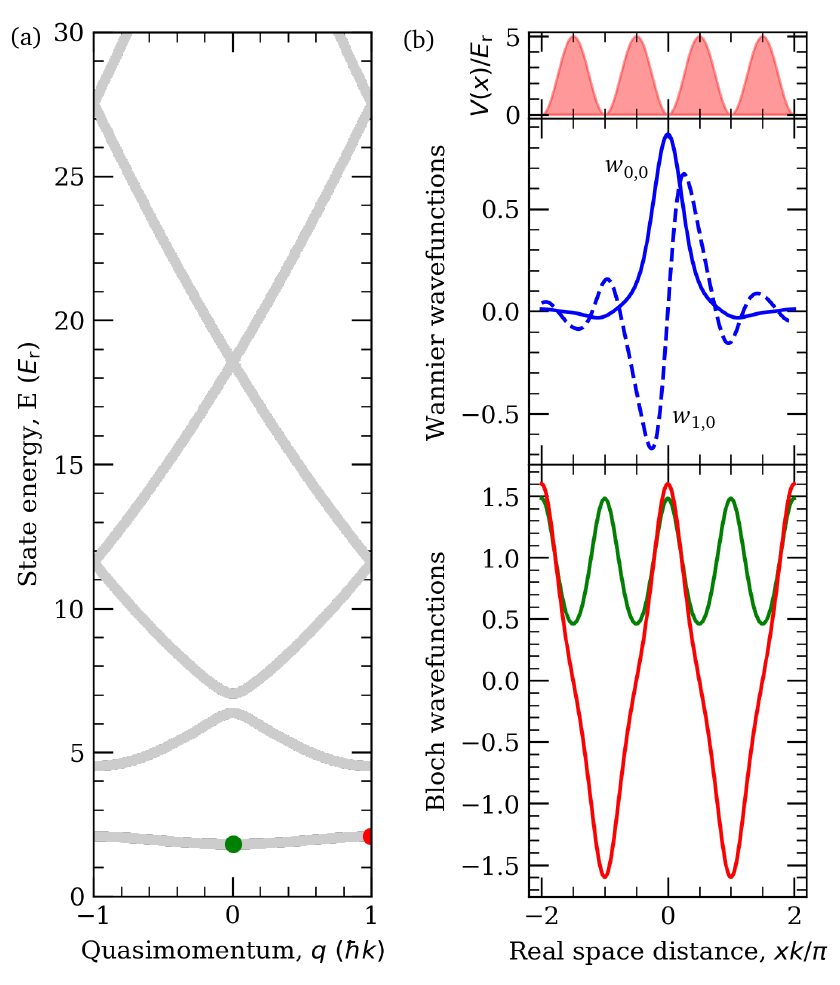}
\end{center}
\vspace{-0.3in}
\caption{\label{fig:bloch_wannier}
Example calculations for Rb atoms trapped in an optical standing wave $V(x)$ with wavelength $\lambda = 1064/2$~nm and depth 5~$E_\mathrm{R}$. (a) Bloch band diagram showing energy as a function of quasimomentum $q$. (b) Lower panel: Corresponding Bloch wave functions for $q=0$ (green) and $q=\hbar k$ (red). Middle panel: Wannier functions for the ground state and first excited Bloch band at $R=0$. Upper panel: optical lattice potential $V(x)$.}

\end{figure}

Atoms in optical lattices are important in many areas of science and technology including atomic clocks and gravimeters, quantum gas microscopes and quantum simulators. A laser standing wave gives rise, through the AC Stark shift
(Sec.~\ref{sec:AC_Stark}),
to a spatially periodic potential for the atoms. As is well-known from solid state physics, such a periodic potential can also be considered in reciprocal space. In momentum space ($k-$basis), a potential with spatial period $\lambda/2$ couples free particle states that
are separated by an integer multiple of $\delta k = 4\pi / \lambda = 2k \equiv k_{\rm l}$
where $\lambda$ is the wavelength
of the optical field, and $k_{\rm l}=2k$ is the lattice momentum.
Therefore, from the Bloch theorem, the eigenfunctions of the Hamiltonian (called Bloch states or Bloch wave functions in this context) can be parametrized by a quasimomentum $q$ ($|q|<k$), and these eigenfunctions are of the form $\exp(iqr)$ times a periodic function of $r$ whose period is the same as that of the lattice. The Hamiltonian can thus be diagonalised on a discrete basis of free-particle states
$\{e^{iqr}, e^{i(q+k_{\rm l})r}, e^{i(q+2k_{\rm l})r},\ldots\}$
separated by an integer
multiple of the lattice momentum in momentum space.
Plotting the resulting energy levels for different
values of the quasimomentum gives Bloch bands~[Fig.~\ref{fig:bloch_wannier}(a)].

Both Bloch states and Bloch bands can be  calculated using
\texttt{OpticalLattice1D}. For example, the following code produces a Bloch band diagram for rubidium atoms trapped in an optical lattice formed by a  1064~nm standing wave with a maximal depth of 5~$E_\mathrm{R}$,
using a basis which includes states up to $q+35 k_{\rm l}$ in momentum ($E_\mathrm{R}=h^2/(2m \lambda^2)$ is the recoil energy):
\begin{lstlisting}
atom = Rubidium87()
trapWavelength = 1064e-9
lattice = OpticalLattice1D(atom, trapWavelength)

lattice.defineBasis(35)
qMomentum = np.linspace(-1,1,100)
lattice.diagonalise(5, qMomentum)
fig = lattice.plotLevelDiagram()
plt.show()
\end{lstlisting}

We can enumerate Bloch bands in order of
their increasing energies, starting with the index 0 for the ground state. The
corresponding Bloch wave functions can be obtained by specifying the quasimomentum and the Bloch band index, as in the following example:
\begin{lstlisting}
# lattice depth = 40 recoil energies
# quasi momentum = + 0 k
# blochBandIndex = 0
f = lattice.BlochWavefunction(40, +0.0, 0)
\end{lstlisting}

The Bloch wave functions are delocalised across the lattice
sites~[Fig.~\ref{fig:bloch_wannier}(b) bottom panel]. For many calculations and a more intuitive mapping to atomic physics experiments, it is
convenient to switch to a basis of localised functions, namely Wannier functions. For each
Bloch band, the Wannier functions $w_{i,R}(x)$ are defined (up to a normalisation factor)
as a complete orthogonal set of functions localised at
lattice points defined by a lattice vector $R$:
\begin{equation}
w_{i,R}(x) \propto \sum_q e^{-iqR} b_{i,q}(x),
\end{equation}
where the sum goes over all values of the quasimomentum $q$ and $b_{i,q}$ are the Bloch
wave functions for a given Bloch band index $i$ and quasimomentum $q$.
Values of the Wannier function can be obtained after diagonalisation of the interaction
potential Hamiltonian for which we defined Bloch band index to be saved
by setting \texttt{saveBandIndex} keyword argument in \texttt{diagonalise} method.
We can then call \texttt{getWannierFunction} to obtain values of the
Wannier function in a given Bloch band~[Fig.~\ref{fig:bloch_wannier}(b) middle panel]. For example:
\begin{lstlisting}
atom = Rubidium87()
trapWavelength = 1064e-9
lattice = OpticalLattice1D(atom, trapWavelength)
lattice.defineBasis(35)
qMomentum = np.linspace(-1,1,100)
lattice.diagonalise(Vlat, qMomentum,
    saveBandIndex=0)
print(lattice.getWannierFunction(x, latticeIndex = 0))
\end{lstlisting}
Note that \texttt{saveBandIndex} selects the band index,
and \texttt{latticeIndex} sets the localisation  of the function at the site with
the corresponding index.
The Wannier functions returned by the program should be normalised on the
relevant lattice by the user.

\subsection{Dynamic polarisabilities and magic wavelengths}\label{sec:AC_Stark}

The dynamic (AC) polarisability $\alpha(\omega)$ of an atom
exposed to an oscillating electric field of angular frequency $\omega$
can be expressed as the sum of a contribution from the polarisability of the valence electron(s) 
$\alpha_{\rm v}$  
and the core polarisability  $\alpha_{\rm c}$.
The valence polarisability often dominates.  For an electron in state
$|a\rangle$ with total angular momentum $J$ and projection $m_J$ the valence polarisability for linearly polarised light can be written as~\cite{Mitroy2010}
\begin{equation}
\alpha_{\rm v}(\omega) = \alpha_0(\omega) +
\frac{3m_J^2 - J(J+1)}{J(2J-1)}\, \alpha_2(\omega) + \alpha_0^{(\mathrm{cont.})}(\omega);
\end{equation}
that is as the sum of a scalar polarisability
\begin{equation}
\alpha_0(\omega) = \frac{2}{3(2J+1)}\sum_{\textrm{states }|b\rangle} 
\frac{|\langle {b}\, ||\, er\, ||\, {a}\rangle|^2 \,
(E_{b}-E_{a})}{ (E_{b}-E_{a})^2 -(\hbar\omega)^2},
\end{equation}
and a tensor polarisability
\begin{align}
\alpha_2(&\omega) = 4\, \left( \frac{5J(2J-1)}{6(J + 1)(2 J + 1)(2 J + 3)} \right)^{1/2} \nonumber \\
& \;\times \sum_{\textrm{states }|b\rangle}
(-1)^{J+J_{b}} \begin{Bmatrix}
J & 1 & J_{b} \\
1 & J & 2
\end{Bmatrix}
\, \frac{|\langle {b}\, || \, er\, ||\, {a}\rangle|^2 \,
(E_{b}-E_{a})}{ (E_{b}-E_{a})^2 -(\hbar\omega)^2}.
\end{align}
Here $\langle {b} \, ||\, er\, ||\, {a}\rangle$ are
reduced dipole matrix elements and the summation runs over all the bound
states $|b\rangle$, with total angular momentum $j_b$, dipole-coupled to the state $|a\rangle$ of interest. Finally, there is also a term
$\alpha_0^{\mathrm{(cont.)}}$ contributed by the continuum of unbound states (this contribution
will be discussed in more detail below).

\begin{figure}[!b]
\begin{center}
\includegraphics[width=\linewidth]{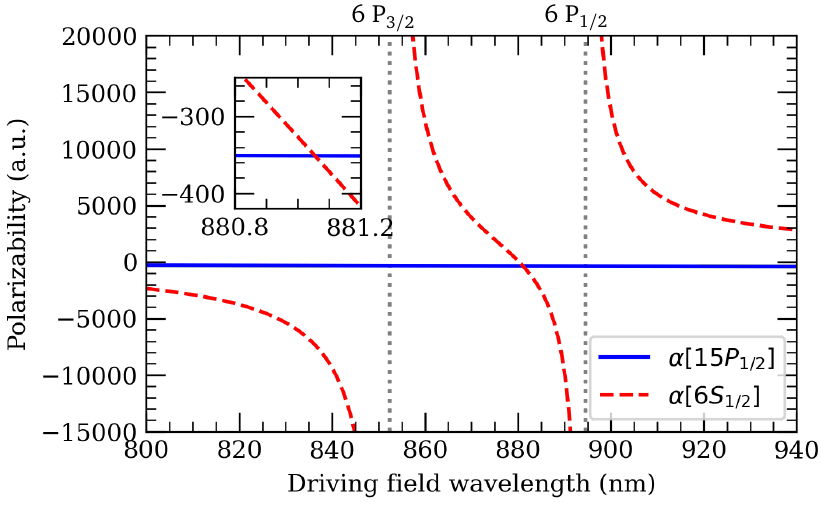}
\end{center}
\vspace{-0.3in}
\caption{\label{fig:polarisability}
Example calculation of $\alpha(\omega)$ for the caesium 6~S$_{1/2}$ (red dashed) and 15~P$_{1/2}$ (blue solid) states. Resonant transitions to other states are marked
with vertical dotted lines. The inset show a region around 881 nm where the polarisabilities are equal, corresponding to a magic wavelength.}
\end{figure}

For example, the scalar and tensor polarisabilities of the caesium
100p~$^2$P$_{1/2}$
state in the AC field given by a 1100~nm wavelength optical trapping laser can be obtained using the following code
(in this calculation, the sum over intermediate states $b$ is limited to all the bound states with $n \leq 115$, including the ground state and the low lying excited states):
\begin{lstlisting}
atom = Caesium()
n = 100
calc = DynamicPolarizability(atom, n, 1, 1.5)
calc.defineBasis(atom.groundStateN, n+15)
alpha0, alpha1, alpha2, alphaC, alphaP, closestState = calc.getPolarizability(1100e-9, units="SI")
\end{lstlisting}
\texttt{closestState} saves the state whose transition frequency 
is closest to that of the driving field. 

In addition to  \texttt{alpha0} and  \texttt{alpha2} the method
\texttt{getPolarizability} also returns the  core polarizability, $\alpha_{\rm c}$
(\texttt{alphaC}), and the ponderomotive polarizability, $\alpha_{\rm p}$ (\texttt{alphaP}). The core polarizability is approximated by its static value (saved in \texttt{atom.alphaC}) which is appropriate when the driving laser is far
from resonance with core transitions. The ponderomotive polarizability can be linked to the contribution from the continuum of unbound states $\alpha_0^{\mathrm{(cont.)}}$ (which is currently not calculated by ARC). Close to a bound-state resonance, the contributions from the bound intermediate states dominate $\alpha_{\rm v}(\omega)$. 
However, for Rydberg states one is often in the opposite limit far from bound-state resonances. Away from strong resonances, the total scalar polarisability
$\alpha_0+\alpha_0^\mathrm{(cont.)}$
then approaches the free-electron ponderomotive polarisability $\alpha_P$ associated with the
time-averaged motion of a free electron in an oscillating electric field
[$\alpha_P = e^2 / (2 h m_e \omega^2)$, where $\omega$
is frequency of driving field and $e$ and $m_e$ are electron charge and mass
respectively]. This result is obtained by applying the limit where the Keppler frequency of the electron orbit around the core is much smaller than $\omega$ as is
typical for Rydberg states~\cite{Dutta2000}, and applying a Born-Oppenheimer
approximation to separate the fast quiver motion
of the loosely bound electron driven by the electric field component and the slower relative
motion of electron around the ionic core.

The code  also returns \texttt{alpha1} which is a vector
polarisability relevant in driving with non-linearly
polarised light, see, e.g., Ref.~\cite{LeKien2013}. This feature is not exploited  in the
current implementation of \texttt{DynamicPolarizability},
which focuses on the simplest case of driving under linearly polarized light;
however it provides a path for future support of
other polarizations. Other future extenstions could include adding bound-state resonances to the core polarizability and an extended treatment of the continuum contribution near the ionization threshold.

\texttt{DynamicPolarizability} calculations can involve dipole matrix elements between low-lying
states or between low-lying states and highly excited states. For low-lying
states dipole matrix elements calculated both with the semiclassical approximation
and the  model potential method become less accurate. That is why 
ARC uses literature values for these states, where available. Users should check the availability of literature values in the corresponding files~(Sec.~\ref{sec:local_data}). If values for
low-lying transitions are available, the expected accuracy can be of
the order of $\sim1\%$, otherwise the accuracy is limited to $\sim 10\%$  for typical
input parameters. Currently, there are significant compiled literature
sources for dipole matrix elements in alkali atoms, but the data for divalent atoms is more
scarce.

For quantum state control in atoms in optical traps, it is often desirable
to find optical trap parameters at which the AC induced shifts of two
states, i.e., the polarisabilities of these two states, is exactly the same.
That happens at so called ``magic wavelengths'' for optical trapping field.
For example, to find such wavelengths for the 15p~$^2$P$_{1/2}~m_J=1/2$ and
 6s~$^2$S$_{1/2}~m_J=1/2$ states of caesium
we plot the polarisability of these two states in the same plot (see Fig.~\ref{fig:polarisability}). The magic wavelengths
are then given by the intersections where the the two polarisabilities are equal:
\begin{lstlisting}
atom = Caesium()
n = 15
calcP = DynamicPolarizability(atom, n, 1, 0.5)
calcP.defineBasis(atom.groundStateN, n+15)

calcS = DynamicPolarizability(atom, 6, 0, 0.5)
calcS.defineBasis(atom.groundStateN, 25)

wavelengthList = np.linspace(800, 940,1000) * 1e-9  # 800-940 nm

ax = calcP.plotPolarizability(wavelengthList, debugOutput=True, units="au")
calcS.plotPolarizability(wavelengthList, debugOutput=True, units="au", addToPlotAxis=ax, line="r--")
plt.show()
\end{lstlisting}
Resonances are indicated by vertical lines (as shown in Fig.~\ref{fig:polarisability}). The states responsible for these resonances and the corresponding resonance wavelengths are also printed when the flag \texttt{debugOutput=True} is set on.

\section{Implementation notes}

\subsection{Handling of the spin quantum number}

To allow the use of both alkali atoms and divalent atoms, most of the methods have
a new keyword argument \texttt{s} specifying the spin quantum number The default
value of this keyword argument, \texttt{s=0.5}, is appropriate for alkali atoms. As such, \texttt{s} does not
have to be specified for these atoms, making the  API of ARC 3.0 completely backwards compatible with
ARC 1.0. For divalent atoms, users should specify whether they are working
with singlet or triplet states by setting the corresponding spin quantum number to
\texttt{s=0} or \texttt{s=1} respectively. To state that the second atom has a
different spin in \texttt{PairStateInteractions},
the optional keyword argument \texttt{s2} should be set explicitly.

\subsection{Use of fitted quantum defects for divalent atoms}
The available experimental energy level data is much sparser and
less precise for divalent atoms than for the alkalis. Inaccuracies arising from experimental errors may be reduced by using energies derived from a fit of the data to the Rydberg-Ritz formula over a range of principal quantum numbers (see Appendix \ref{app:defects}), rather than using the experimental energies directly. However, in many cases the calculations will require energies outside the range of values of $n$ used in the fitting to the Rydberg-Ritz formula.  Using this formula is normally justified for values of $n$ above this range, but might be invalid for values of $n$ below this range (e.g., because the Rydberg-Ritz formula does not apply to the ground state and low excited states, or because the energy levels do not vary smoothly and monotonically with $n$ due to perturbers). Using the experimental energies is thus often preferable for small values of $n$. 

For clarity, we have tabulated the range of values of $n$ over which the quantum defects have been fitted using a 
dictionary variable \texttt{defectFittingRange}  indexed by terms.
For each term the dictionary returns the smallest and largest principal quantum
numbers defining the range of principal quantum numbers used in the calculation of the Rydberg-Ritz coefficients provided in ARC. For example, the following code will do this for the $^1$F$_3$ term in Calcium-40:
\begin{lstlisting}
atom = Calcium40()
term = '1F3'
if term in atom.defectFittingRange:
    fitRange = atom.defectFittingRange[term]
    print("Min n = %d" % fitRange[0])
    print("Max n = %d" % fitRange[1])
\end{lstlisting}
For principal quantum numbers \emph{above} the fitted range, the energies are calculated by extrapolation of the Rydberg-Ritz formula to outside this range. No extrapolation is done
for principal quantum numbers \emph{below} the fitted range. Instead, ARC tries to use tabulated energies if they exist --- either energies provided by the user or the energies provided by ARC, which are literature values as described in the online documentation \cite{arc3_documentation}.
In the case of S, P or D series, the calculation is aborted and a value error raised, explaining the problem, if the program requires a missing tabulated energy. For $L > 2$, the calculation is not aborted but a value of zero is assumed for the quantum defect.
These tabulated energies must be stored in the local data
directory (Sec.~\ref{sec:local_data}) under a file name
defined for each atom in the \texttt{levelDataFromNIST} variable.
For example for strontium they are stored in \texttt{sr\_level\_data.csv}.
The \texttt{energyLevelsExtrapolated} variable
of the used atom will be set to \texttt{True} if the calculation uses energies obtained by extrapolation of the Rydberg-Ritz formula to outside the range of values of $n$ used in the fitting.

\subsection{Local data directory and updates to literature values}\label{sec:local_data}

Dipole matrix elements from the literature can be added to ARC by setting the \texttt{literatureDMEfilename} variable in the \texttt{DivalentAtom} class to the appropriate file name and ensuring the format of the file matches that specified in the documentation.

When ARC 3.0 is first used, a local hidden directory named \texttt{.arc-data} is
created  in the user's home folder. The results of calculations of, e.g, dipole matrix
elements are saved in human-readable files in this folder, forming a look-up table that can be used to speed up future calculations.
There is an option to add dipole matrix elements from the literature (or obtained
using other calculation methods) by modifying  the corresponding files. For example,
to add a new literature value for a $^{88}$Sr dipole matrix element, the user should
modify the file \texttt{strontium\_literature\_dme.csv} as per the table header. This can
be conveniently done by loading the appropriate .csv file in any spreadsheet program, as long as the file format is preserved
in terms of separators and header comments. To change
other atomic properties, like quantum defects, users should make their 
own subclasses which inherit classes defined in the ARC.
An example is given in the following code,
which could be used to update the quantum defects for calcium:
\begin{lstlisting}
class MyCalcium40(Calcium40):
    quantumDefect = ...
    # write updated quantum defects
    # list in order specified in documentation
    dipoleMatrixElementFile = "my_ca_dme.npy"
\end{lstlisting}
Once defined, \texttt{MyCalcium40} can be used instead of \texttt{Calcium40} in all
the ARC calculations. 
To ensure the program does not use old cached values, new names
of caching files should be provided when redefining atomic species in this way. Also, note
that if a new version of arc-data is released (changing the
 version number in \texttt{version.txt} in the data directory), the data entered manually in the \texttt{.arc-data} folder may  be
overwritten automatically if no new file names had been specified.

We encourage users to submit new experimental data and parameters for use by the community via a pull request  on
the ARC GitHub page~\cite{arcGitHub}.

\subsection{Looking under the hood with debugOutput=True}

As for ARC 1.0, setting the keyword argument \texttt{debugOutput=True}
results in verbose output, which may be useful for checking basis states and other intermediate
results. Additionally, many methods have a \texttt{progressOutput=True} option which
can be used for tracking the progress of the calculations.

\section{Outlook}

In summary, we have presented ARC~3.0, a major new release of the ARC Python library  that extends the library to divalent atoms and adds a number of new methods of general interest in Rydberg physics and beyond.
We believe this common code base and consistent interface for many different
atomic properties can speed up the development of new applications and lower knowledge barriers, e.g, in quantum technologies based on neutral atoms.  
The library also offers rich possibilities of advanced
educational projects for students in atomic and quantum physics.

Future improvements of ARC could include the addition of the calculation of wave functions for divalent atoms
and additional methods for the 
accurate calculation of low-lying wave functions~\cite{Kozlov2015},
support for multi-channel quantum defect
theory calculations~\cite{Vaillant2014},
and the calculation of scattering properties.
New experimental data can be straightforwardly added to the existing base. ARC is a community-oriented open source package, and the authors welcome contributions of new core data or algorithms to the main library as well as contributions of more specialized codes to \texttt{arc.advanced}.

\section*{Acknowledgements}
We thank Tsz-Chun Tsui and Trey Porto for tabulating the ytterbium data. We also thank Hei Yin Andrea Kam for the use of her code to check the calculations of strontium dipole matrix elements, Paul Huillery for help with Stark map calculations, and Ifan Hughes for suggesting the bootstrap method.  N.~\v{S}. is supported by the H2020 Marie Sk{\l}odowska-Curie Actions (COQUDDE, H2020-MSCA-IF-2017 grant agreement No.786702). The project was supported by the EPSRC Platform grant EP/R002061/1 and Standard grant EP/R035482/1. We also acknowledge funding from the project EMPIR-USOQS (EMPIR projects are co-funded by the European Union's Horizon2020 research and innovation programme and the EMPIR Participating States).

\appendix

\section{The single active electron approximation}\label{sec:single_e_approx}
Atomic structure calculations for Rydberg states are simplified by the rapid scaling of the size of the wave function of the Rydberg electron with principal quantum number $n$. In atoms with a single valence electron, the calculation of energy levels and wave functions is achieved by solving the Schr\"{o}dinger equation with a modified ``model'' potential that accounts for the screening effects of the closed-shell core. Each Rydberg series labelled by the quantum numbers $L,S$ and $J$ is characterized by a quantum defect $\delta{n}$ that quantifies the deviation of each energy level from its hydrogenoic equivalent. This method works well for alkali atoms and is used by \texttt{alkali\_atom\_functions}. 

For divalent atoms, the situation is complicated by two  effects. The first is that the ionic core is no longer a closed shell, since it contains the remaining valence electron. The core is therefore more strongly polarized by the Rydberg electron, leading to an additional $n$ dependence in the quantum defects. The second effect is the existence of compact states where both electrons are excited (e.g., a 5p$^2$ configuration), known as perturbers.
These perturber states exhibit strong interelectronic correlation effects and lead to perturbations of nearby Rydberg levels. Even away from the energy of a perturber, Rydberg states may  acquire a small admixture of doubly excited states \cite{Gallagher1994}.

Nevertheless, under many circumstances the properties of Rydberg states in divalent atoms can be described under a single active electron approximation \cite{Vaillant2012}. Under this approximation, Russell-Saunders $L-S$ coupling is assumed to hold, and $L,S$ and $J$ are regarded as good quantum numbers. The effects of core polarizability and perturbers are partially included via the energy dependence of the quantum defects. Calculations can then be made in a similar way to those performed in alkali atoms, with the appropriate generalization of the angular momentum algebra to integer spin. Previous work has shown that this treatment gives good agreement with experiment for quantities that depend on the long-range part of the wave function, such as the DC Stark effect  \cite{Millen2011spectro,Bowden2017}. 

The interaction between two divalent Rydberg atoms is also dominated by the long-range character of the wave function. Therefore a single active electron approach may be used here also \cite{Vaillant2012}. A study of the effect of perturbers on the interactions \cite{Vaillant2015} demonstrated that the single-electron treatment is valid to a high degree of accuracy (<2\%) except for Rydberg states in the immediate vicinity of perturbers. 

Note that other observables that depend on the short-range properties of the wave function, such as the coupling to low-lying states (eg radiative lifetimes), are significantly modified by the presence of even small amounts of perturbing states, and are not well treated in the single active electron approximation \cite{Vaillant2014}. Here other methods such as multi-channel quantum defect theory (MQDT) \cite{Vaillant2014} that explicitly include the effects of perturbers must be used.

\section{Atomic Data for Divalent Atoms}\label{app:defects}
\setcounter{table}{0} 
As outlined in section \ref{sec:semiclassical}, calculations involving divalent atoms rely on fitted values for the quantum defects. In the case of Ca, no new data was available, and so values compiled in \cite{Vaillant2012,Vaillant2014a} are used. For Sr and Yb, we provide a new analysis of the available data for Sr and Yb that takes into account recent improvements in the spectroscopic data.

Experimental energy levels were fitted to the modified Rydberg-Ritz formula 
\begin{equation}
    E(n) = I_{\mathrm{a}} - \frac{Ry_\mathrm{a}}{\left(n-\left[\delta_0 + \displaystyle{ \frac{\delta_2}{(n-\delta_0)^2} + \frac{\delta_4}{(n-\delta_0)^4}}\right]\,\right)^2}
    \label{app:RR}
\end{equation}
using $\delta_0$, $\delta_2$ and $\delta_4$ as free parameters. The ionization energy $I_{\mathrm{a}}$ was constrained to the values obtained from the analysis of the best available spectroscopic data. $Ry_\mathrm{a}$ is the atom-specific Rydberg constant: $Ry_\mathrm{a} = R_\infty\, {m_\mathrm{a}}/({m_\mathrm{a}+m_\mathrm{e}})$, where $R_\infty$ is the Rydberg constant, and $m_\mathrm{a}$ and $m_\mathrm{e}$ are the mass of the considered isotope and the electron respectively. The fine-structure splitting of the ionization threshold was neglected. A least-squares fitting method was used, implemented via the  $ \texttt{curve\_fit}$ function from the \texttt{scipy} Python package.

The results, along with references to the experimental dataset are provided in Table \ref{tab:defect_table}

\
\begin{table*}
  \caption{Tabulated values of  the Rydberg-Ritz parameters for $^{88}\mathrm{Sr}$ and $^{174}\mathrm{Yb}$. Here $I_\mathrm{Sr} = 45\,932.2002\ \mathrm{cm}^{-1}$ and $Ry_\mathrm{Sr}= 109\,736.631\ \mathrm{cm}^{-1}$ \cite{Couturier2019} and $I_\mathrm{Yb} = 50443.07041\ \mathrm{cm}^{-1}$ and $Ry_\mathrm{Yb}= 109\,736.96959\ \mathrm{cm}^{-1}$ \cite{Lehec2018}. Lower ($\mathrm{cl}_-$) and upper ($\mathrm{cl}_+$) confidence limits are defined such that 68\% of the values obtained by bootstrap resampling and fitting procedure lie within the interval $[\delta_j-\mathrm{cl}_-,\delta_j+\mathrm{cl}_+ ]$
   }
    \centering
    \begin{tabular}{cc | lll|  lll | lll | c | l }
    \hline \mbox{}&&&&&&&&&&&&\\[-3mm]
        \multicolumn{2}{c}{Series}     & \multicolumn{3}{c}{$\delta_0$} & \multicolumn{3}{c}{$\delta_2$} & \multicolumn{3}{c}{$\delta_4$}  & Range & Refs \\
        \hline
        \mbox{}&&&&&&&&&&&&\\[-3mm]
        {}&{}& $\delta_0$&$\mathrm{cl}_-$& $\mathrm{cl}_+$ & $\delta_2$&$\mathrm{cl}_-$& $\mathrm{cl}_+$& $\delta_4$& $\mathrm{cl}_-$& $\mathrm{cl}_+$  & $n_1,n_2$ & {} \\[1mm]
        $^{88}\mathrm{Sr}$  &$^1$S$_0$~ & 3.26912 & $2\times10^{-5}$ & $6\times10^{-5}$ & -0.178 & 0.023 &0.009 & 3.5 & 0.7 &3.0 &  14, 34 & \cite{Rubbmark1978, Beigang1982a}\\
        &$^3$S$_1$~ & 3.370773 & $3\times10^{-6}$ & $4\times10^{-6}$ & 0.420 &0.003 &0.001& -0.4&0.1&0.3& 15, 50 & \cite{Sansonetti2010, Couturier2019} \\
        &$^1$P$_1$~ & 2.724& $0.002$ & 0.006 & -3.4 & 1.3 & 0.3 & -220 & 10  & 70 & 10, 29 & \cite{Rubbmark1978, Baig1998, Sansonetti2010} \\
        &$^3$P$_0$~ & 2.8867 & $5\times10^{-4}$&  $9\times10^{-4}$ & 0.43 & 0.10 & 0.05 & -1.8 & 0.6& 3 & 8, 15 & \cite{Esherick1977, Sansonetti2010} \\
        &$^3$P$_1$~ & 2.8826 & $5\times10^{-4}$& $9\times10^{-4}$ & 0.39 & 0.14 & 0.04 & -1.1 & 0.7 & 5 & 8, 21 & \cite{Armstrong1979, Sansonetti2010}\\
        &$^3$P$_2$~ & 2.882 & 0.003 & 0.002 & -2.5& 1.8 & 1.8& 100 & 300 & 300& 19, 41 & \cite{Armstrong1979, Sansonetti2010}\\
        &$^1$D$_2$~ & 2.3847& $4 \times 10^{-4}$ & $6 \times 10^{-4}$ & -42.0 &0.6 & 0.4 & -620 & 60 & 140 & 20, 50 & \cite{Esherick1977a, Beigang1982a, Couturier2019} \\
        &$^3$D$_1$~ & 2.67524 & $5 \times 10^{-5}$ & $7 \times 10^{-5}$ & -13.23 & 0.15 & 0.08 & -4420 & 40& 90& 28, 50 & \cite{Sansonetti2010, Couturier2019} \\
        &$^3$D$_2~$ & 2.66149& $6 \times 10^{-5}$  & $10 \times 10^{-5}$  & -16.9 & 0.3 & 0.2 & -6630 & 50 & 120 & 28, 50 & \cite{Esherick1977a, Couturier2019} \\
        &$^3$D$_3$~ & 2.66 & 0.04 &0.03 & -70 & 30 & 40  & -14000 & 9000 & 6000&  20, 37 & \cite{Sansonetti2010, Beigang1983} \\
        &$^1$F$_3$~ & 0.0909 & $5 \times 10^{-4}$ & $18 \times 10^{-4}$ & -2.44 & 0.64 & 0.12 & 62 & 9& 60&  10, 28 & \cite{Rubbmark1978, Sansonetti2010}\\
        &$^3$F$_2$~ & 0.120 & 0.001 & 0.004 & -2.2 & 1.4 & 0.3 & 100 & 20 & 120 & 10, 24 & \cite{Rubbmark1978, Sansonetti2010} \\
        &$^3$F$_3$~ & 0.119 &0.003 & 0.004 & -2.0 & 1.3 & 0.6& 100 & 40 & 110 & 10, 24 & \cite{Rubbmark1978, Sansonetti2010}\\
        &$^3$F$_4$~ & 0.120 & 0.002 & 0.003& -2.4 & 0.9 & 0.4 & 120 & 30 & 70 & 10, 28 & \cite{Rubbmark1978, Sansonetti2010}\\[3mm]
        $^{174}\mathrm{Yb}$ &$^1$S$_0$~ & 4.27837 & $3 \times 10^{-5}$ & $4 \times 10^{-5}$ & -5.61 & 0.09 & 0.07 & -260 & 40 & 50 &  34, 80 &\cite{Aymar1980,Lehec2018}\\
        &$^1$P$_1$~ & 3.95343 & $2\times 10^{-5}$ & $3\times 10^{-5}$ & -10.58 & 0.06 & 0.05  & 730 & 30 & 50 & 35, 54 &\cite{Lehec2018,Maeda1992}\\
        &$^1$D$_2$~ & 2.71301 & $9\times 10^{-5}$ & $7\times 10^{-5}$ &  -0.9 & 0.3 & 0.4 & -600 & 400 & 300 & 40, 80& \cite{Lehec2018}\\
        &$^3$D$_2$~ & 2.74860 & $6\times 10^{-5}$ & $5 \times 10^{-5}$ & 0.01 & 0.15 & 0.16 & -100 & 120 & 110 & 35, 80& \cite{Aymar1980,Lehec2018}\\[1mm]
        \hline
    \end{tabular}
    \label{tab:defect_table}
\end{table*}
The choice of the range of $n$ used for each series was a compromise between maximising the number of data points, and reducing the effect of series perturbations not described by equation \ref{app:RR} and experimental systematic uncertainties. Particular care should be taken in extrapolating to values of $n$ below the stated range, where the fits are often poor.

The uncertainties on the Rydberg-Ritz parameters are 68\% confidence limits obtained using a ``bootstrap" method based on resampling with replacement \cite{Hughes2010}, such that each confidence interval includes 68\% of the results falling above the quoted value of the corresponding parameter as well as 68\% of the results falling below it. For each series the fitted range was resampled 150 times. The asymmetry of the confidence limits reflects the asymmetric dependence of the value of the energy on the quantum defects encapsulated in Eq.~(\ref{app:RR}), as well as the limitations of the experimental data. Note that correlations in the uncertainties are expected to be strong and are not explicitly considered here; users seeking to set rigorous error bounds on derived quantities should take this into account.

\section{ARC 3.0 API}\label{app:function_listing}
\setcounter{table}{0} 
The tables~\ref{tab:alkalifunc} to~\ref{tab:wigner} list the APIs of ARC~3.0.
All old methods have been amended to
allow handling of divalent atoms with two possible spin states. Additionally, methods
that are newly introduced, or have significant new functionality, are marked with a
$\blacklozenge$ . Since for divalent atoms the calculation of electronic wave functions
in the model potential approach is not implemented, some methods are available only when working
with \texttt{AlkaliAtom} instances, and these are marked by a $\circ$.

\begin{table*}[h!]
\caption{\label{tab:alkalifunc}Class and function listing of the  \texttt{alkali\_atom\_functions} module.}
\centering
\footnotesize
\begin{tabular}{l l}
\hline 
Name (parameters) & Short description\\
\hline
\verb|AlkaliAtom|([preferQuantumDefects, cpp\_numerov])  & Implements general calculations for alkali atoms (see Table~\ref{tab:alkaliatom})\\
\verb|NumerovBack|(innerLimit, outerLimit, kfun, ...) & Full Python implementation of Numerov integration\\
\verb|saveCalculation|(calculation, fileName) & Saves calculation for future use\\
\verb|loadSavedCalculation|(fileName) &  	Loads previously saved calculation\\
\verb|printStateString|(n, l, j) &	Returns state spectroscopic label for $|n,l,j\rangle$ \\
\hline 
\end{tabular}
\end{table*}

\begin{table*}[ht]
\caption{\label{tab:alkali_and_divalent_atom} Methods and function listing of the \texttt{alkali\_atom\_functions.AlkaliAtom} and \texttt{alkali\_atom\_functions.DivalentAtom} classes. \texttt{getRadialCoupling} now uses semiclassical calculations (see~\ref{sec:semiclassical}) for \texttt{DivalentAtom}. The typical relative uncertainties are obtained from comparison to measured values. \label{tab:alkaliatom}}
\centering
\footnotesize
\begin{tabular}{l l c}
\hline  
Name (parameters) & Short description (units) & Typical rel. accuracy\\
\hline
\verb|getDipoleMatrixElement|(n1, l1, ...) &	Reduced dipole matrix element ($a_0 e$)& $ \sim 10^{-2}$\\
\verb|getDipoleMatrixElementHFS|(n1, l1, ...) &	$\blacklozenge$ Hyperfine-structure resolved transitions $\langle n_1\ell_1 j_1f_1m{f1}|er| n_2\ell_2j_2f_2 m_{f2}\rangle$ ($a_0 e$)& $ \sim 10^{-2}$\\
\verb|getTransitionWavelength|(n1, l1, ...) &	Calculated transition wavelength in vacuum (m)&  $ \sim 10^{-6}$\\
\verb|getTransitionFrequency|(n1, l1, ...) & Calculated transition frequency (Hz)  &  $  \sim 10^{-6}$\\
\verb|getRabiFrequency|(n1, l1, j1, mj1, ...) &Returns a Rabi frequency (angular, i.e. $\Omega = 2\pi\times \nu$) for resonant excitation  & \\
& with a specified laser beam in the center of TEM00 mode (rad~s$^{-1}$)& $ \sim 10^{-2}$\\
\verb|getRabiFrequency2|(n1, l1, j1, mj1, ...) &Returns a Rabi frequency (angular, i.e. $\Omega = 2\pi\times \nu$) for resonant excitation & \\
& with a specified electric field driving amplitude (rad~s$^{-1}$)& $ \sim 10^{-2}$\\
\verb|getStateLifetime|(n, l, j[, ...]) & Returns the lifetime of the state (s)& $ \sim 10^{-2}$\\
\verb|getTransitionRate|(n1, l1, j1, n2, ...) & Transition rate due to coupling to vacuum modes (black body included) (s$^{-1}$)& $ \sim 10^{-2}$\\
\verb|getReducedMatrixElementJ_asymmetric|(n1, ...) & Reduced matrix element in $J$ basis, defined in asymmetric notation ($a_0 e$)& $ \sim 10^{-2}$\\
\verb|getReducedMatrixElementJ|(n1, l1, ...) & Reduced matrix element in $J$ basis, symmetric notation ($a_0 e$)& $ \sim 10^{-2}$\\
\verb|getReducedMatrixElementL|(n1, l1, ...) & Reduced matrix element in $L$ basis, symmetric notation ($a_0 e$)& $ \sim 10^{-2}$\\
\verb|getRadialMatrixElement|(n1, l1, ...) & Radial part of the dipole matrix element ($a_0 e$)& $ \sim 10^{-2}$\\
\verb|getQuadrupoleMatrixElement|(n1, ...) & Radial part of the quadrupole matrix element ($a_0^2 e$)& $ \sim 10^{-2}$\\
\verb|getPressure|(temperature) & Vapour pressure at a given temperature (Pa)& $\sim (1-5) \cdot 10^{-2}$\\
\verb|getNumberDensity|(temperature) & Atom number density at a given temperature (m$^{-3}$)& $\sim (1-5) \cdot 10^{-2}$\\
\verb|getAverageInteratomicSpacing|(...) & Returns the average inter-atomic spacing in the atomic vapour (m)& $\sim (1-5) \cdot 10^{-2}$\\
\verb|corePotential|(l, r) & $\circ$ core potential felt by the valence electron (a.u)&\\
\verb|effectiveCharge|(l, r) & $\circ$ effective charge of the core felt by the valence electron (a.u)&\\
\verb|potential|(l, s, j, r) & $\circ$ potential(l, s, j, r) (a.u)&\\
\verb|radialWavefunction|(l, s, j, ...) & $\circ$ Radial part of the electron wave function&\\
\verb|getEnergy|(n, l, j) &  	Energy of the level relative to the ionisation level (eV)& $  \sim 10^{-6}$\\
\verb|getZeemanEnergyShift|(l, j, mj, ...) & $\blacklozenge$  Returns the linear (paramagnetic) Zeeman shift. (J)& \\
\verb|getQuantumDefect|(n, l, j) & Quantum defect of the level.&\\
\verb|getC6term|(n, l, j, n1, l1, j1, ...) & $C_6$ interaction term for the given two pair-states ($h~ \times~$Hz~m$^6$)&\\
\verb|getC3term|(n, l, j, n1, l1, j1, ...) & $C_3$ interaction term for the given two pair-states ($h~ \times~$Hz~m$^3$)&\\
\verb|getEnergyDefect|(n, l, j, n1, l1, ...) & Energy defect for the given two pair-states, $E(|rr\rangle) - E(|r'r''\rangle)$ (eV)&\\
\verb|getEnergyDefect2|(n, l, j, nn, ll, ...) & Energy defect for the given two pair-states, $E(|r_1r_2\rangle) - E(|r'r''\rangle)$ (eV)&\\
\verb|updateDipoleMatrixElementsFile|() & Updates the file with pre-calculated dipole matrix elements&\\
\verb|getRadialCoupling|(n, l, j, n1, l1, j1) & $\blacklozenge$ Returns the radial part of the coupling between two states (dipole,&\\
& quadrupole) ($a_0 e$ or $a_0^2 e$)& $ \sim 10^{-2}$\\
\verb|getAverageSpeed|(temperature) & Average one-dimensional speed at a given temperature (m/s)&\\
\verb|getLiteratureDME|(n1, l1, j1, n2, ...) & Returns literature information on requested transition&\\
\hline  
\end{tabular}
\end{table*}

\begin{table*}[ht]
\caption{Class listing of the  \texttt{alkali\_atom\_data} module. All these classes inherit properties of  \texttt{alkali\_atom\_functions.AlkaliAtom} from Table~\ref{tab:alkaliatom}.}
\centering
\footnotesize
\begin{tabular}{l l}
\hline 
Name (parameters) & Short description\\
\hline
\verb|Hydrogen|([preferQuantumDefects, cpp\_numerov]) &	Properties of hydrogen atoms\\
\verb|Lithium6|([preferQuantumDefects, cpp\_numerov]) &	Properties of lithium-6 atoms\\
\verb|Lithium7|([preferQuantumDefects, cpp\_numerov]) &	Properties of lithium-7 atoms\\
\verb|Sodium|([preferQuantumDefects, cpp\_numerov]) &	Properties of sodium-23 atoms\\
\verb|Potassium39|([preferQuantumDefects, cpp\_numerov]) & 	Properties of potassium-39 atoms; alias \verb|Potassium(...)|\\
\verb|Potassium40|([preferQuantumDefects, cpp\_numerov]) & 	Properties of potassium-40 atoms\\
\verb|Potassium41|([preferQuantumDefects, cpp\_numerov]) & 	Properties of potassium-41 atoms\\
\verb|Rubidium85|([preferQuantumDefects, cpp\_numerov]) &	Properties of rubidium-85 atoms; alias \verb|Rubidium(...)|\\
\verb|Rubidium87|([preferQuantumDefects, cpp\_numerov]) &	Properties of rubidium-87 atoms\\
\verb|Caesium|([preferQuantumDefects, cpp\_numerov]) &	Properties of caesium-133 atoms;
alias \verb|Cesium(...)| \\
\hline 
\end{tabular}
\end{table*}

\begin{table*}[ht]
\caption{Class listing of the \texttt{divalent\_atom\_data} module. All these classes inherit the properties of  \texttt{divalent\_atom\_functions.DivalentAtom} from Table~\ref{tab:alkali_and_divalent_atom}.}
\centering
\footnotesize
\begin{tabular}{l l}
\hline  
Name (parameters) & Short description\\
\hline
\verb|Strontium88|([preferQuantumDefects]) &	$\blacklozenge$ Properties of strontium-88 atoms\\
\verb|Calcium40|([preferQuantumDefects]) &	$\blacklozenge$ Properties of calcium-40 atoms\\
\verb|Ytterbium174|([preferQuantumDefects]) & $\blacklozenge$	Properties of ytterbium-174 atoms\\
\hline  
\end{tabular}
\end{table*}

\begin{table*}[ht]
\caption{Method listing of the  \texttt{calculations\_atom\_single.LevelPlot}(atomType) class.}
\centering
\footnotesize
\begin{tabular}{l l}
\hline
Name (parameters) & Short description\\
\hline
\verb|makeLevels|(nFrom, nTo, lFrom, lTo) &	Constructs an energy level diagram in a given range\\
\verb|drawLevels|() &	Draws a level diagram plot\\
\verb|showPlot|() &	Shows a level diagram plot\\
\hline 
\end{tabular}
\end{table*}

\begin{table*}[ht]
\caption{Method listing of the \texttt{calculations\_atom\_single.StarkMap}(atom) class.}
\centering
\footnotesize
\begin{tabular}{l l}
\hline 
Name (parameters) & Short description\\
\hline
\verb|defineBasis|(n, l, j, mj, nMin, ...) &	Initializes a basis of states around the state of interest\\
\verb|diagonalise|(eFieldList[, ...]) &	Finds atom eigenstates in a given electric field\\
\verb|plotLevelDiagram|([units, ...]) &	Makes a plot of a Stark map of energy levels\\
\verb|showPlot|([interactive]) &	Shows plot made by \verb|plotLevelDiagram| \\
\verb|savePlot|([filename]) &	Saves plot made by \verb|plotLevelDiagram|\\
\verb|exportData|(fileBase[, exportFormat]) &	Exports \verb|StarkMap| calculation data\\
\verb|getPolarizability|([maxField, ...]) &	Returns the polarizability of the state (MHz~cm$^2$/V$^2$) \\
\verb|getState|(state, electricField, ...) & $\blacklozenge$ Returns the state composition for the state with a largest contribution of a target state in given E-field  \\
\hline 
\end{tabular}
\end{table*}

\begin{table*}[ht]
\caption{Method listing of the \texttt{calculations\_atom\_single.Wavefunction}() class}
\centering
\footnotesize
\begin{tabular}{l l}
\hline 
Name (parameters) & Short description\\
\hline
\verb|getRtimesPsiSpherical|(theta, phi, r) & $\blacklozenge$ $\circ$ Calculates the list of $r\cdot \psi_{m_s}(\theta,\phi,r)$ for all possible $m_s$\\
\verb|getRtimesPsi|(x, y, z) &	$\blacklozenge$ $\circ$ Calculates the list of $r \cdot \psi_{m_s}(x,y,z)$ for all possible $m_s$\\
\verb|getPsi|(x, y, z) & $\blacklozenge$ $\circ$ Calculates the list of $\psi_{m_s}(x,y,z)$ for all possible $m_s$\\
\verb|getRtimesPsiSquaredInPlane|([...]) & $\blacklozenge$ $\circ$ Calculates $|r \cdot \psi|^2$ on a mesh in a given plane \\
\verb|plot2D|([plane, pointsPerAxis, ...]) & $\blacklozenge$ $\circ$ 2D colour plot of $|r \cdot \psi|^2$ wave function in a requested plane \\
\verb|plot3D|([plane, pointsPerAxis, ...]) & $\blacklozenge$ $\circ$ 3D colour surface plot of $|r \cdot \psi|^2$ wave function in a requested plane \\
\hline 
\end{tabular}
\end{table*}

\begin{table*}[ht]
\caption{Method listing of the  \texttt{calculations\_atom\_single.AtomSurfaceVdW}() class.}
\centering
\footnotesize
\begin{tabular}{l l}
\hline 
Name (parameters) & Short description\\
\hline
\verb|getC3contribution|(n1, l1, j1, ...) & $\blacklozenge$ Contribution to $C_3$ of the
$|n_1,\ell_1,j_1\rangle$ state due to dipole coupling to the $|n_2,\ell_2,j_2\rangle$ state. (J$\cdot$m$^3$ )\\
\verb|getStateC3|(n, l, j, ... [, s, ...]) & $\blacklozenge$ van der Waals atom-surface interaction coefficient for a given state (J$\cdot$m$^3$ ) \\
\hline 
\end{tabular}
\end{table*}

\begin{table*}[ht]
\caption{Method listing of the \texttt{calculations\_atom\_single.OpticalLattice1D}() class.}
\centering
\footnotesize
\begin{tabular}{l l}
\hline 
Name (parameters) & Short description\\
\hline
\verb|getRecoilEnergy|() &  $\blacklozenge$ Recoil energy for atoms in given optical lattice (J)\\
\verb|getTrappingFrequency|(...) & $\blacklozenge$ Atom's trapping frequency for a given trap depth (Hz) \\
\verb|defineBasis|([Limit]) & $\blacklozenge$ Define the basis for a Bloch band calculation\\
\verb|diagonalise|(... [, ... ]) & $\blacklozenge$ Calculates energy levels (Bloch bands) for a given list of quasimomenta\\
\verb|plotLevelDiagram|() & $\blacklozenge$ Plots energy level diagram (Bloch bands) \\
\verb|BlochWavefunction|(... ) & $\blacklozenge$ Bloch wave function as a \emph{function} of 1D coordinate\\
\verb|getWannierFunction|(x[, ...]) & $\blacklozenge$ Gives value of a Wannier function \\
\hline 
\end{tabular}
\end{table*}

\begin{table*}[ht]
\caption{Method listing of the \texttt{calculations\_atom\_single.DynamicPolarizability}() class}
\centering
\footnotesize
\begin{tabular}{l l}
\hline 
Name (parameters) & Short description\\
\hline
\verb|defineBasis|(nMin, nMax) &  $\blacklozenge$ Defines basis for the calculation of the dynamic polarizability\\
\verb|getPolarizability|(...) & $\blacklozenge$ Calculates the scalar and tensor polarizabilities \\
\verb|plotPolarizability|(...) & $\blacklozenge$ Plots the polarisability for a range of wavelengths\\
\hline 
\end{tabular}
\end{table*}

\begin{table*}[ht]
\caption{Class listing of the  \texttt{materials} module. Each class inherits the abstract
class \texttt{OpticalMaterial}() that implements the literature input of data,
while \texttt{getN} in each specific class has arguments
according to the requirements of the material (e.g. multiple axes of reflection etc).}
\centering
\footnotesize
\begin{tabular}{l l}
\hline 
Name (parameters) & Short description\\
\hline
\verb|Air()|(...) & $\blacklozenge$ Air at normal conditions as an optical material \\
\verb|Sapphire|(...) & $\blacklozenge$ Sapphire as an optical material\\
\hline 
\end{tabular}
\end{table*}

\begin{table*}[ht]
\caption{Method listing of the $\blacklozenge$  \texttt{calculations\_atom\_pairstate.PairStateInteractions}(atom, n, l, j, nn, ll, jj, m1, m2, interactionsUpTo=1) class that calculates the Rydberg level diagram (a ``spaghetti diagram") for the given pair-state. The $\blacklozenge$ symbols for this class indicate
significant additions of functionality: \texttt{PairStateInteractions} now supports
arbitrary \emph{inter-species} pair-state calculations, and \texttt{getC6perturbatively}
supports degenerate perturbation calculations.}
\centering
\footnotesize
\begin{tabular}{l l}
\hline 
Name (parameters) & Short description\\
\hline
\verb|defineBasis|(theta, ...) &	Finds the relevant states in the vicinity of the given pair-state\\
\verb|getC6perturbatively|(...) & $\blacklozenge$ 	Calculates $C_6$ coefficients from second order [degenerate] perturbation theory (GHz~$\mu$m$^6$)\\
\verb|getLeRoyRadius|() & $\circ$ Returns the Le Roy radius for the initial pair-state ($\mu$m)\\
\verb|diagonalise|(rangeR, ...) &	Finds eigenstates in atom pair basis\\
\verb|plotLevelDiagram|([...]) &	Plots a pair-state level diagram\\
\verb|showPlot|([interactive]) &	Shows the level diagram printed by \verb|plotLevelDiagram|\\
\verb|exportData|(fileBase[, ...]) &	Exports \verb|PairStateInteractions| calculation data\\
\verb|getC6fromLevelDiagram|(...) & 	Finds the $C_6$ coefficient for the original pair-state (GHz~$\mu$m$^6$)\\
\verb|getC3fromLevelDiagram|(...) & 	Finds the $C_3$ coefficient for the original pair-state (GHz~$\mu$m$^3$)\\
\verb|getVdwFromLevelDiagram|(...) &	Finds the $r_{\rm vdW}$ coefficient for the original pair-state ($\mu$m)\\
\hline 
\end{tabular}
\end{table*}

\begin{table*}[ht]
\caption{Method listing of the  \texttt{calculations\_atom\_pairstate.StarkMapResonances}(atom1, state1, atom2, state2) class that calculates pair-state Stark maps for finding resonances.}
\centering
\footnotesize
\begin{tabular}{l l}
\hline 
Name (parameters) & Short description\\
\hline
\verb|findResonances|(nMin, ...) &	Finds near-resonant dipole-coupled pair-states\\
\verb|showPlot|([interactive]) &	Plots a Stark map for the initial state and its dipole-coupled resonances\\
\hline 
\end{tabular}
\end{table*}

\begin{table*}[ht]
\caption{\label{tab:wigner}Function and class listing of the  \texttt{wigner} module providing support for angular element calculations}
\centering
\footnotesize
\begin{tabular}{l l}
\hline 
Name (parameters) & Short description\\
\hline
\verb|CG|(j1,m1, ...) & returns a Clebsch-Gordan (CG) coefficient\\
\verb|Wigner3j|(j1,j2, ...) &	returns a Wigner 3j-coefficient\\
\verb|Wigner6j|(j1,j2,...) &	returns a Wigner 6j-coefficent\\
\verb|wignerDmatrix|(theta,phi) & Class for obtaining Wigner D-matrices\\
\hline 
\end{tabular}
\end{table*}

\section*{References}

\end{document}